\documentclass[pre,twocolumn]{revtex4-1}
\usepackage{bm}
\usepackage{graphicx}
\usepackage{amsmath}
\usepackage{amssymb}
\usepackage{amscd}
\usepackage{epstopdf}
\usepackage{verbatim}
\usepackage[usenames]{color}
\usepackage{array}
\usepackage{xspace}
\usepackage{xstring}
\usepackage[normalem]{ulem}

\newcommand{\ie}{\textit{i.e.}\xspace}
\newcommand{\eg}{\textit{e.g.}\xspace}
\newcommand{\Dr}{D_\text{r}}
\newcommand{\vel}{\mathrm{v}} 

\definecolor{Blue}{rgb}{0,0,0.8}
\definecolor{BlueB}{rgb}{0,0,0.5}
\definecolor{Red}{rgb}{0.8,0,0}
\definecolor{RedB}{rgb}{0.6,0,0}
\definecolor{Green}{rgb}{0,0.5,0}
\definecolor{GreenB}{rgb}{0,0.8,0}
\definecolor{Purple}{rgb}{1,0,1}

\renewcommand{\vec}[1]{{\bf #1}}
\newcommand{\vecu}[1]{\hat{\bf #1}}

\newcommand{\vecug}[1]{\hat{\bm #1}}

\newcommand{\rhopsi}{\rho^\psi}

\newcommand{\flux}{J} 
\newcommand{\region}[1]{#1} 

\begin{document}

\title{
Dynamics and density distribution of strongly confined \\ noninteracting nonaligning self-propelled particles in a nonconvex boundary
}
\author{Yaouen Fily, Aparna Baskaran, Michael F. Hagan}
\affiliation{Martin Fisher School of Physics, Brandeis University, Waltham, MA 02453, USA}
\date{\today}


\begin{abstract}
We study the dynamics of non-aligning, non-interacting self-propelled particles confined to a box in two dimensions.
In the strong confinement limit, when the persistence length of the active particles is much larger than the size of the box, particles stay on the boundary and align with the local boundary normal.
It is then possible to derive the steady-state density on the boundary for arbitrary box shapes.
In non-convex boxes, the non-uniqueness of the boundary normal results in hysteretic dynamics and the density is non-local, \ie it depends on the global geometry of the box.
These findings establish a general connection between the geometry of a confining box and the behavior of an ideal active gas it confines, thus providing a powerful tool to understand and design such confinements.
\end{abstract}

\maketitle

\section{Introduction}

Active systems are non-equilibrium systems whose constituent units consume energy to generate motion or mechanical forces. Originally inspired by biology, \eg bacterial colonies~\cite{Dombrowski2004,Peruani2012}, healing tissues~\cite{Poujade2007,Trepat2009}, or flocking animals~\cite{Ballerini2008}, the field now encompasses a variety of artificial systems that share this ability to inject energy at the microscopic level and emulate the unique properties of their biological counterparts, from flocking to spontaneous aggregation~\cite{Palacci2010a,Paxton2004, Hong2007,Jiang2010,Volpe2011,Thutupalli2011,Bricard2013,Narayan2007,Kudrolli2008,Deseigne2010}.
Beyond biomimetism, the study of active matter has led to new applications not found in nature, such as bacteria-powered micro-gears~\cite{Angelani2009,DiLeonardo2010,Sokolov2010}.

One distinctive yet relatively unexplored property of active systems is their sensitivity to boundary effects.
Striking macroscopic effects may be obtained by patterning confining walls on the micro-scale, as exemplified by the rectification phenomenon~\cite{Galajda2007,Wan2008,Tailleur2009,Angelani2011,Ghosh2013,Ai2013}.
More generally, any real-world system must have boundaries, and understanding their role is paramount to designing active matter based devices.
Whether the boundaries are only present by necessity or designed as an integral component of an active system, it is important to note that boundary effects are not merely size effects: the exact shape of the boundary is crucial.
However, most existing studies are only concerned with one among a handful of specific geometries~\cite{Tailleur2009,Nash2010,Elgeti2013,Lee2013, Mallory2014,Kaiser2012,Kaiser2013,Wioland2013,Kantsler2013, Guidobaldi2014,Yang2014,Camley2014,Ray2014,Lushi2014,Kaiser2014}, and little is known about how the shape of a boundary affects the behavior of the active system it confines.

In this paper, we focus on non-aligning self-propelled particles, a model that has recently attracted attention as a minimal model for self-propelled matter~\cite{Fily2012,Redner2013,Bialke2013,Cates2013,Stenhammar2013,Fily2014, Stenhammar2014,Wittkowski2014,Mallory2014}.
In particular, we neglect alignment interactions such as those that would arise from hydrodynamic couplings in a fluid environment.
Our results thus apply to systems in which such coupling torques are weak.
Furthermore, we restrict ourselves to the ``ideal active gas'' limit in which particles interact with the wall, but not with each other~\cite{Mallory2014}.
We recently showed, for such a system, an analytic relationship between the density and pressure of the active gas and the shape of the box for a general class of box shapes~\cite{Fily2014a}.
In the strong confinement regime, where the persistence length of the active particles is much larger than the size of the box, and when the box is convex, we showed that particles never leave the boundary and always align their self-propulsion direction with the local boundary normal.
Furthermore, the density and the pressure on the boundary are proportional to the local boundary curvature.
It is then possible to predict the density and pressure
on the boundary of any convex box, regardless of the details of its shape.
However, existing applications suggest that active devices are most effective when their boundaries have both convex and concave regions.

In this paper, we extend the theoretical framework introduced in Ref.~\cite{Fily2014a} to the case of non-convex boxes.
The presence of concave regions is a significant complication, as it implies that the same normal is found at multiple locations on the boundary, leading to multi-stability and hysteresis. Furthermore, particles within concave regions undergo complex, accelerated dynamics that sometimes launches them off the wall.
Nonetheless, we demonstrate that in the strong confinement regime: (i) this complex dynamics can be understood in terms of non-local ``jumps'',
(ii) the density of particles within concave regions vanishes and (iii) it is possible to predict the density everywhere on the boundary.  We present a general algorithm to obtain this relationship and we test our predictions against the results of molecular dynamics simulations in a family of boxes with both concave and convex regions.
Finally, we discuss the role of interactions and the limits of the ideal gas approximation.

The paper is arranged as follows.
Section~\ref{model} introduces the model.
Section~\ref{wall_dynamics} explores the particle dynamics on the boundary and shows how the accelerated dynamics over concave regions can be recast as instantaneous jumps between disparate convex regions (see also appendices~\ref{ap:jump_typology} and~\ref{ap:jump_duration}).
Section~\ref{quasi_density} presents a theory for the density on the boundary in the strong confinement regime, and shows how to obtain the steady-state density.
Section~\ref{simulations} presents the results of molecular dynamics simulations and compares them against the predictions of sections~\ref{wall_dynamics} and~\ref{quasi_density}.
Section~\ref{discussion} discusses  the scope of our model and the role of convexity in confined active gases.

\section{Model}
\label{model}

We consider a collection of confined, overdamped, self-propelled particles in two dimensions. Each particle is characterized by its position $\vec{r}$ and orientation $\vecug{\nu}=\cos\theta\,\vecu{x} + \sin\theta\,\vecu{y}$. The dynamics obeys the following equations of motion:
\begin{align}
\dot{\vec{r}} = \vel_0 \vecug{\nu} + \mu \vec{F}_\text{w}
\, , \quad
\dot{\theta} = \xi(t)
\label{eq:eom1}
\end{align}
where $\vel_0$ is the self-propulsion speed, $\mu$ is the mobility, $\xi$ is a white Gaussian noise with zero mean and correlations $\langle\xi(t)\xi(t')\rangle=2\Dr\delta(t-t')$, and over-dots indicate time derivatives.
The medium on which the self-propulsion force is exerted is treated as a momentum sink; in particular, we neglect hydrodynamic interactions.
The confining walls are hard and frictionless; whenever the velocity of a particle is such that it would drive the particle into the wall, its component normal to the wall is cancelled by the wall force. More precisely, the wall force $\vec{F}_\text{w}$ is zero if the particle is not at the wall, or if it is at the wall but pointing away from it; otherwise it is equal to
$-\left(\frac{\vel_0}{\mu}\vecug{\nu}_i\cdot\vecu{n}\right)\vecu{n}$, where
$\vecu{n}=\cos\psi\,\vecu{x} + \sin\psi\,\vecu{y}$ is the local normal to the wall pointing outwards~\footnote{With this choice of potential, the dynamics does not depend on the value of the mobility $\mu$ which is only kept for dimensional consistency.}.
This is the simplest choice of wall potential consistent with overdamped dynamics.
It neglects alignment terms that can arise when particles are anisotropic or experience hydrodynamic effects, and thusis appropriate when the re-orientation rate induced by such torques is slow in comparison to particle translation (this point is discussed further in section~\ref{discussion}).
Finally, since particles are non-interacting, we may restrict the discussion to a single particle.

\begin{figure} 
\centering
\includegraphics[width=0.45\linewidth]{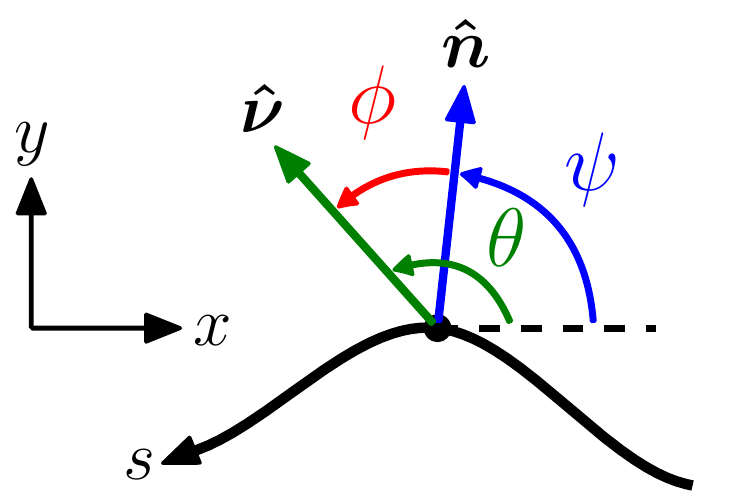}
\caption{ (Color online)
Notations for a particle at the wall.
The particle is characterized by its arclength $s$ along the wall and its orientation $\vecug{\nu}=\cos\theta\,\vecu{x} + \sin\theta\,\vecu{y}$.
The local normal to the wall is $\vecu{n}=\cos\psi\,\vecu{x} + \sin\psi\,\vecu{y}$.
}
\label{fig:sketches}
\end{figure}

When a particle is at the wall, its configuration is characterized by its arclength $s\in [0,L)$ along the boundary, where $L$ is the box perimeter, and its orientation 
relative to the local boundary normal $\phi=\theta-\psi$ (see Fig.~\ref{fig:sketches}).

There are two important lengths scales in the system: the active persistence length $\vel_0/\Dr$, \ie the typical distance a free (unconfined) particle travels before its orientation decorrelates, and the global size of the confining box.
However, for a boundary with nonuniform curvature, the variations in the local radius of curvature lead to additional length scales. Which one is most relevant depends strongly on the geometry of the box, as discussed in Ref.~\cite{Fily2014a} and in the rest of the paper.

The regime we study in this paper is the \emph{strong confinement regime}, obtained when the persistence length $v_0/\Dr$ is much larger than the size of the box.
It is obtained at large self-propulsion, small angular noise or small box size, and
we will often refer to it as the \emph{small angular noise regime}, or simply the \emph{small noise regime} (the angular noise is the only noise in our model).

\section{Dynamics at the wall}
\label{wall_dynamics}

We first look at the dynamics of a single particle moving along the wall. In particular, we explore the fundamental difference between convex and concave regions and show how the latter cause fast jumps and bi-stability in the low noise regime.
To this end, we project the equations of motion~\eqref{eq:eom1} onto the tangent to the wall:
\begin{align}
\dot{s} = \vel_0 \sin\phi
\, , \quad
\dot{\phi} = \xi(t) - \frac{\vel_0}{R(s)}\sin\phi
\label{eq:eom2}
\end{align}
where $s$ is the arclength along the wall, $\phi=\theta-\psi$ is the angle between the particle's orientation $\vecug{\nu}$ and the boundary normal $\vecu{n}$ (see Fig.~\ref{fig:sketches}), and $R(s)=ds/d\psi$ is the local radius of curvature.
Eqs.~\eqref{eq:eom2} remain valid as long as $|\phi|\le\pi/2$; as soon as $|\phi|>\pi/2$, the particle leaves the boundary.

\subsection{Dynamics at zero angular noise}
\label{zero_dynamics}

\begin{figure} 
\centering
\includegraphics[width=0.85\linewidth]{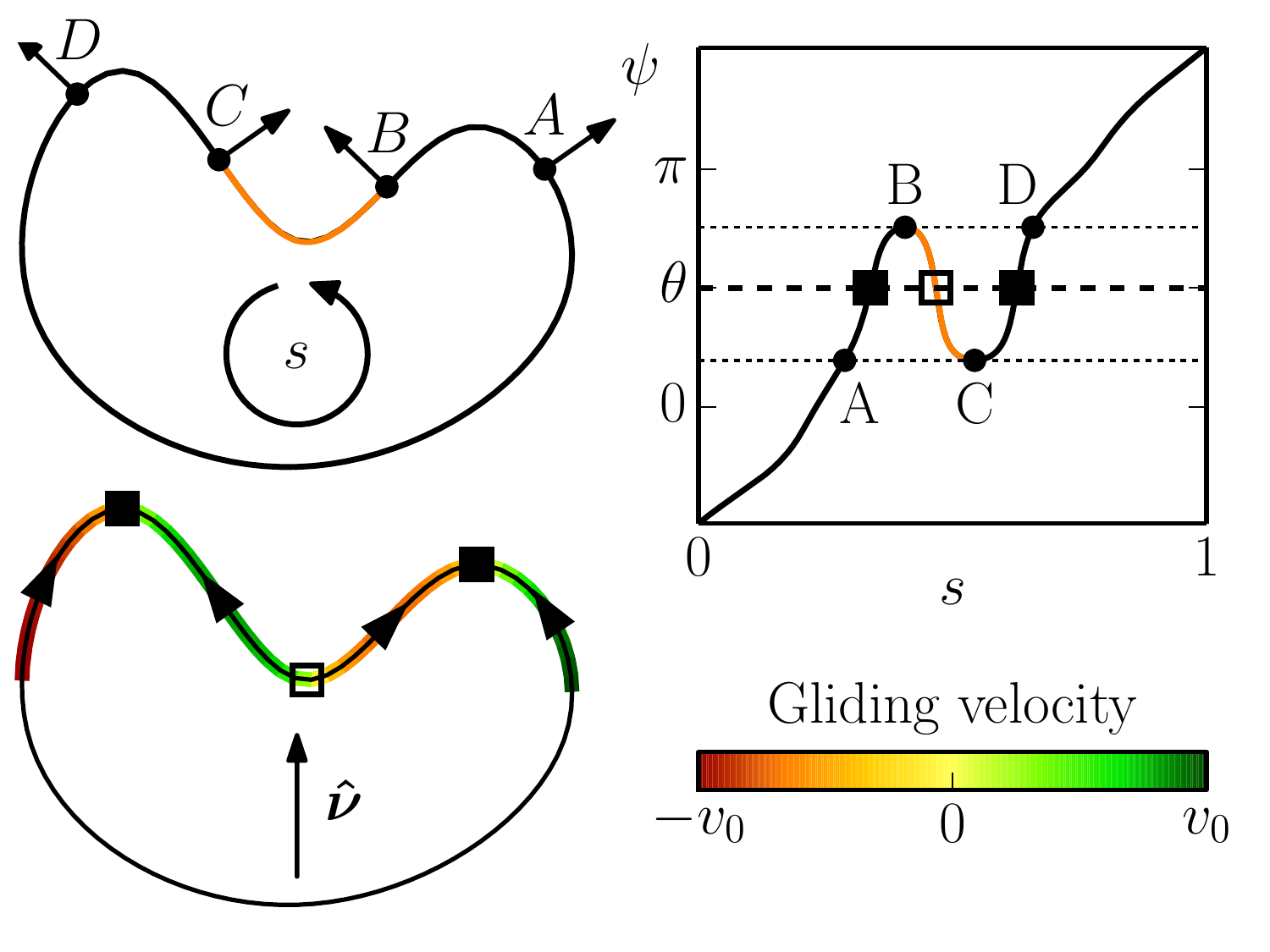}
\caption{ (Color online)
Dynamics in a confining box in the absence of angular noise.
Top left:
Real space representation. The concave region (between $B$ and $C$) is shown in orange (grey). The arclength $s$ is counted counter-clockwise.
The points $A$ and $D$ have the same normal (shown as a straight arrow) as $C$ and $B$, respectively.
Top right:
Normal angle $\psi$ as a function of arclength $s$.
The dynamics of a particle with constant orientation $\theta$ is controlled by the distance $\phi=\theta-\psi$ between the curve and the horizontal dashed line.
The fixed points as shown as filled (stable) and empty (unstable) squares.
Bottom left:
Gliding velocity $\dot{s}=v_0\sin\phi$
for a particle with orientation $\vecug{\nu}$ (shown inside the box).
The orientation and fixed points are the same as those shown in the top right panel.
Arrow heads indicate the gliding direction.
The colorbar is shown on the right.
}
\label{fig:gliding_speed}
\end{figure}

In the absence of noise ($\Dr=0$), the orientation $\vecug{\nu}=\cos\theta\vecu{x}+\sin\theta\vecu{y}$ of the particle is constant and its gliding velocity only depends on its location
(see Fig.~\ref{fig:gliding_speed}):
\begin{align}
\dot{s} = v_0 \sin(\theta-\psi(s))
\label{eq:eom4}
\end{align}
When $|\theta-\psi|>\pi/2$ (lower uncolored half of the box in the bottom left panel of Fig.~\ref{fig:gliding_speed}), there is no gliding. Instead, the particle travels in a straight line through the interior of the box until it hits the boundary again.

Locations along the boundary where $\psi(s)=\theta$ (\ie the particle is aligned with the normal) act as fixed points.
They are stable in convex regions ($R(s)>0$) and unstable in concave regions ($R(s)<0$), as can be seen by linearizing Eq.~\eqref{eq:eom4} near the arclength $s_0$ of the fixed point:
\begin{align}
\frac{d}{dt}(s-s_0) = -\frac{v_0}{R(s_0)} (s-s_0)
+ {\cal O}\left((s-s_0)^2\right)
\label{eq:eom5}
\end{align}

Graphically, the fixed points are located at the intersection(s) of the curve $y=\psi(s)$ with the horizontal dashed line $y=\theta$ in the top right panel of Fig.~\ref{fig:gliding_speed}.
In convex boxes, $\psi(s)$ is monotonic and the fixed point corresponding to the orientation $\theta$ is always stable and unique.
Conversely, in the presence of concavity there are multiple locations with the same normal, and multiple fixed points for some values of $\theta$.

\subsection{Quasi-static dynamics}
\label{quasi_dynamics}

We now let the orientation $\theta$ vary slowly. When the rate of change of $\theta$ is small enough, the particle spends most of its time at a fixed point, and only a small fraction of its time
travelling between fixed points.
This \emph{quasi-static} regime is obtained when the angular noise $\Dr$ is small, \ie in the strong confinement regime.
The particle is then confined to the boundary: the fixed point condition $\theta=\psi$ implies that the particle always points toward the boundary, whereas leaving the boundary would require pointing away from it ($|\theta-\psi|>\pi/2$).

In general, a small change $d\theta$ in the orientation $\theta$ causes a small displacement $ds=R(s)d\theta$ of the corresponding fixed point.
In a convex region, the particle relaxes exponentially toward the new fixed point.
The quasi-static regime is then obtained when the corresponding relaxation time $R(s)/v_0$ is much shorter than the reorientation time $\Dr^{-1}$.

In a concave region, on the other hand, an infinitesimal change in the orientation $\theta$ can trigger a large displacement, which we refer to as a \emph{jump}.
Consider the box shown in Fig.~\ref{fig:gliding_speed}, and a particle at point $B$ with orientation $\theta=\psi_B+d\theta$ where $d\theta>0$ is a small perturbation.
Since there is no fixed point with normal angle $\psi_B+d\theta$ in the vicinity of $B$, the particle has to travel to the next convex location with normal angle $\psi_B$, \ie point $D$.
Concretely, the perturbation $d\theta$ sends the particle into the concave region where its gliding speed continuously increases. It only starts to decelerate once it reaches the end $C$ of the concave region, and eventually comes to a stop at point $D$. 
The quasi-static regime is obtained when the jump from $B$ to $D$ is much faster than the reorientation time $\Dr^{-1}$.
In appendix~\ref{ap:jump_duration} we show that this is the case for small angular noise. We also discuss the possibility and implications of particles leaving the boundary during a jump.
Within the quasi-static approximation, however, the details of a jump are irrelevant: it is considered instantaneous, and only its landing point matters.

In summary, the presence of concavity causes non-trivial dynamics in the quasi-static regime.
A particle reaching the end of a convex region experiences an instantaneous jump to a new convex location with the same normal angle.
As a result, the vicinity of a concave region exhibits bi-stability and hysteresis (in Fig.~\ref{fig:gliding_speed}, jumps from $B$ to $D$ and from $C$ to $A$ create an hysteresis loop around $ABCD$).

Finally, these results have important consequences for the steady-state density (see section~\ref{quasi_density}).
First, the instantaneousness of the jumps over concave regions implies that those regions are empty.
Second, the fact that jumps do not stop at the end of the concave region but continue into the next convex region causes non-local density fluxes within the system.
As we show next, the requirement that these fluxes cancel at steady-state enables predicting the density profile everywhere on the boundary.

\section{Quasi-static steady-state density}
\label{quasi_density}

We now use the results of section~\ref{wall_dynamics} to predict the steady-state density of a particle in a box of arbitrary shape in the quasi-static regime.
Our starting point is the assumption that the particle is always at a stable fixed point, \ie at a convex point where the particle's orientation is aligned with the boundary normal ($\theta=\psi$).
This has three important consequences.
First, the particle always points toward the boundary and never leaves it. As a result, the density is zero in the bulk and the problem is effectively limited to the (one-dimensional) boundary.
Second, the particle never visits concave regions, where fixed points are always unstable. The density on the boundary therefore vanishes in those regions.
Third, the normal angle at the location of the particle follows a simple random walk:
$\dot{\psi}=\dot{\theta}=\xi(t)$
where $\xi$ is the same noise as in Eqs.~\eqref{eq:eom1} and~\eqref{eq:eom2}.
Thus, the density on the boundary in $\psi$ space (unit circle) obeys the usual diffusion equation:
\begin{align}
\partial_t \rhopsi = \Dr \partial_\psi^2 \rhopsi
\label{eq:diffusion_total}.
\end{align}
whose steady-state solution is given by $\rhopsi(\psi)=\dfrac{1}{2\pi}$.

\subsection{Role of convexity}

In a convex box, $R(s)$ is positive everywhere and $\psi(s)$ is monotonic.
The density of particles per unit length of boundary is then obtained by making the change of variable $\psi\rightarrow s$. At steady-state, this yields \cite{Fily2014a}
\begin{align}
\rho(s) = \rhopsi(\psi)\, \frac{d\psi}{ds}=\frac{1}{2\pi R(s)}
\end{align}
The density on the boundary is thus proportional to the local boundary curvature.

In a non-convex box, on the other hand, there are multiple locations on the boundary with the same normal angle $\psi$ (see Fig.~\ref{fig:gliding_speed}).
Unlike the density $\rho(s)$, the normal angle density $\rhopsi(\psi)$ does not discriminate between those locations. Thus, $\rho(s)$ cannot be inferred from $\rhopsi(\psi)$ alone for a box with concave boundary regions.

\subsection{Formulating the problem}

To retain all the information contained in $\rho(s)$ while working in normal angle space, where the dynamics is simply diffusive, we number the convex regions $1$ to $n$ and introduce the normal angle density $\rhopsi_i$ in region $i$.
\begin{figure}
\centering
\includegraphics[width=0.95\linewidth]{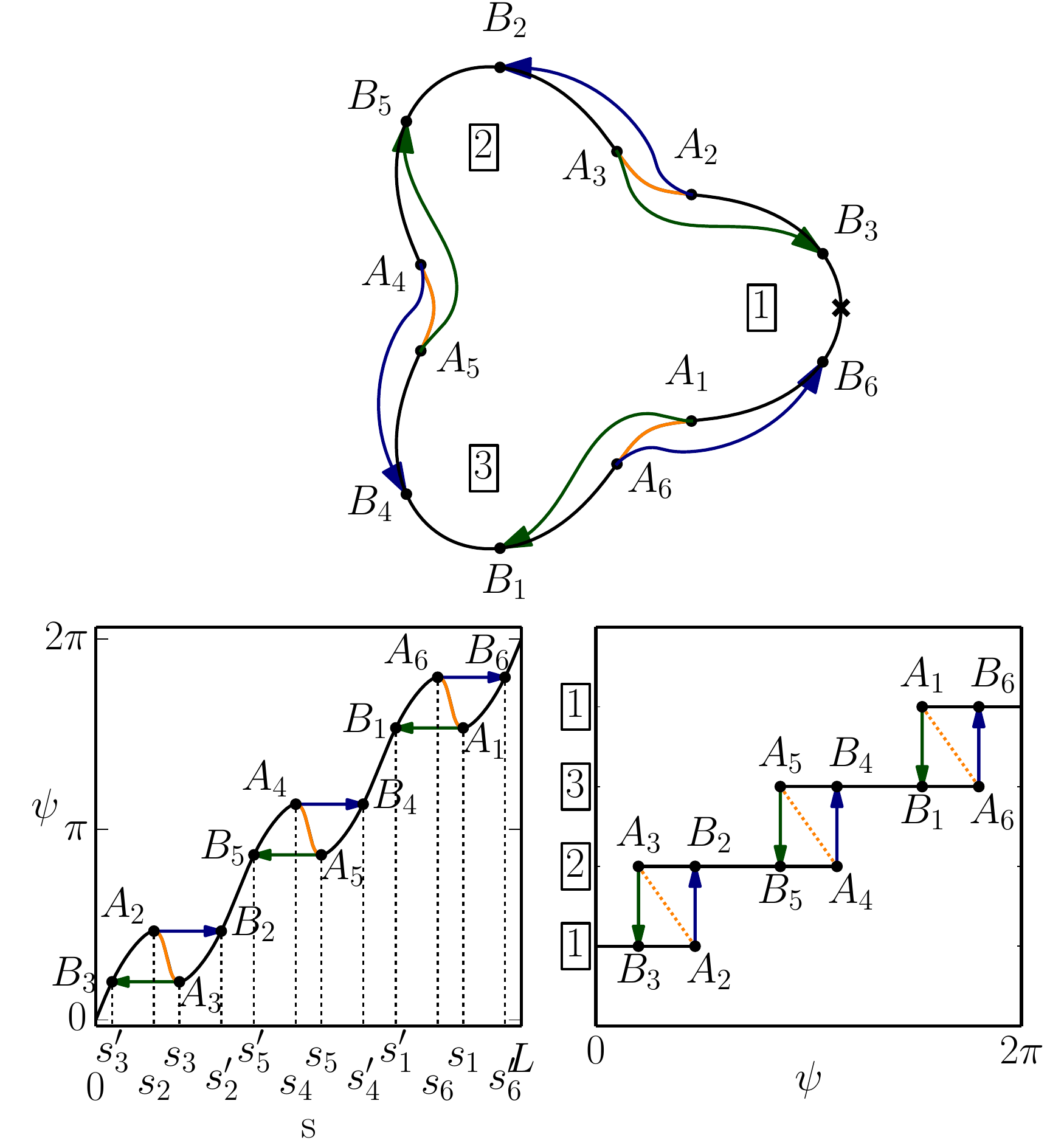}
\caption{ (Color online)
Three representations of a non-convex box showing the concave regions in orange (grey) and the  jumps over the concave regions (curved arrows). Convex regions are indexed by a number $1$ to $3$.
Region $i$ is delimited by the two inflexion points $A_{2i-1}$ and $A_{2i}$.
Each $A_i$ is the starting point of a jump over the neighboring concave region that lands at $B_i$, which has the same normal angle $\psi_i$ as $A_i$. 
Top: Shape of the box. The `x' on the right is the arclength origin, $s=0$.
Bottom left: Normal angle vs. arclength.
Bottom right: Normal angle representation.
The vertical axis labels the convex region corresponding to each interval.
Region \region{1} appears split into two parts due to periodic boundary conditions.
}
\label{fig:topology}
\end{figure}
Region $i$ is delimited by the two inflexion points $A_{2i-1}$ and $A_{2i}$ (see Fig.~\ref{fig:topology}).
Inflexion point $A_i$ has normal angle $\psi_i$ and arclength $s_i$.
Each $\rhopsi_i$ is defined over the entire $[0,2\pi)$ interval but only takes non zero values between $\psi_{2i-1}$ and $\psi_{2i}$, so that we can write $\rhopsi=\sum_i \rhopsi_i$.
The concave regions, where the density is assumed to be zero, are not explicitly described, but manifest themselves through the boundary conditions $\rhopsi_i(\psi_{2i-1})=\rhopsi_i(\psi_{2i})=0$.
Inflexion points act as one-way teleportation devices that send the particle to a new convex location with the same normal angle $\psi$.
These instantaneous jumps often, but not always, connect consecutive regions (see appendix~\ref{ap:jump_typology}).

From the point of view of each $\rhopsi_i$, the start of a jump is a particle sink and its end a particle source. Apart from jumps, the dynamics in a convex region is indistinguishable from that in a convex box and $\rhopsi_i(\psi)$ obeys the same diffusion equation as $\rhopsi(\psi)$.
The result is a set of coupled diffusion equations:
\begin{align}
\partial_t \rhopsi_i = \Dr \partial_\psi^2 \rhopsi_i
+ \sum_k \epsilon_{ik} \flux_k \delta(\psi-\psi_k)
\label{eq:diffusion}
\end{align}
where $\delta$ is the Dirac delta function and the sum is over jumps. 
Jump $k$ occurs at normal angle $\psi_k$ and carries a current $\flux_k>0$ from its starting point $A_k$ to its landing point $B_k$.
$\epsilon_{ik}$ encodes the relationship between jump $k$ and region $i$.
If jump $k$ starts in region $i$, $\epsilon_{ik}=-1$ and the corresponding term in Eq.~\eqref{eq:diffusion} is a sink.
Conversely, if jump $k$ lands in region $i$, $\epsilon_{ik}=1$ and the corresponding term in Eq.~\eqref{eq:diffusion} is a source.
Finally, if jump $k$ does not involve region $i$ then $\epsilon_{ik}=0$; in other words the sum in Eq.~\eqref{eq:diffusion} is restricted to jumps that involve region $i$%
~\footnote{With these notations, it is manifest that summing over $i$ recovers Eq.~\eqref{eq:diffusion_total}, with $\rhopsi(\psi)=\sum_i \rhopsi_i(\psi)$.}.
In order for $\partial_t\rhopsi_i$ to remain finite, each jump must create a discontinuity in $\partial_\psi \rhopsi_i$ proportional to $\flux_k$:
$\partial_\psi \rhopsi_i(\psi_k^+)-\partial_\psi \rhopsi_i(\psi_k^-)=-\flux_k/\Dr$
where the superscripts $\pm$ symbolize one-sided limits. 
The currents $\flux_k$ may then be eliminated from Eq.~\eqref{eq:diffusion} in favor of the densities $\rhopsi_i$.

Once the density in normal angle space is known in every convex region, the linear density on the boundary $\rho(s)$ is obtained using the change of variable $\psi\rightarrow s$, which is monotonic within each convex region:
\begin{align}
\rho(s) =
\begin{cases}
\dfrac{\rhopsi_i(\psi(s))}{R(s)} & $ if $s$ is in convex region $i \\
0 & $ if $s$ is in a concave region$
\end{cases}
\label{eq:unfold}
\end{align}
where $R(s)$ is the radius of curvature.

\subsection{Steady-state}
\label{steady-state}

From the form of Eq.~\eqref{eq:diffusion}, it is clear that the steady-state density in each convex interval 
is piecewise linear, with a change of slope at the location $\psi_k$ of every jump that starts or lands in the interval.
We also require $\rhopsi_i$ to be continuous, and to vanish at the ends of the interval (beyond which the boundary is concave and thus empty):
$\rhopsi_i(\psi_{2i-1})=\rhopsi_i(\psi_{2i})=0$.
As a result, the entire set of density functions $\{\rhopsi_i(\psi)\}_{1\le i\le n}$
is fully determined by its $2n$ values at the location of every jump landing $B_k$ (see Fig.~\ref{fig:topology}).
Let $x_k = \rhopsi_{i(k)}(\psi_k), k\in[1,2n]$ be those unknowns, with $i(k)$ the index of the convex region in which jump $k$ lands.

On the other hand, Eq.~\eqref{eq:diffusion_total} implies that the steady-state total density $\rhopsi=\sum_i \rhopsi_i$ is equal to $1/(2\pi)$.
Using the piecewise linearity of $\rhopsi_i$, $\rhopsi_i(\psi)$ can always be expressed as a linear combination of $x_k$'s. 
Writing $\rhopsi(\psi)=(2\pi)^{-1}$ at $2n$ distinct values of $\psi$ then leads to a $2n\times2n$ linear system that can be solved to obtain the $x_k$'s.
In order for the system not to be degenerate, the $2n$ values of $\psi$ need to be spread across all subregions of all convex regions so that no $x_k$ is left out.
A convenient choice is to use the locations $\psi_k$ of the jumps.

Once the $\rhopsi_i$'s are known, the density per unit length of the boundary is given by Eq.~\eqref{eq:unfold}.

The entire process can be automated, \ie it is possible to write a program that, starting with a parametrization of the boundary, identifies the convex regions and the jumps, writes the linear system, solves for the $x_k$'s, and generates the functions $\rhopsi_i$ and $\rho$.
We used such a program to create most of our figures and analyze our simulation results (section~\ref{simulations}).

Below we go through the method step-by-step in several situations of interest.
In sections~\ref{multiplicity2} and~\ref{multiplicity3}, we consider boxes in which the number of convex locations with the same normal is limited to $2$ (section~\ref{multiplicity2}) and $3$ (section~\ref{multiplicity3}). In both cases, we show that the density can be expressed in a simple form.
In section~\ref{general_case} we give a detailed description of the algorithm that leads to the steady-state density in the general case.

\subsubsection{Multiplicity no greater than $2$}
\label{multiplicity2}

\begin{figure} 
\centering
\includegraphics[width=0.8\linewidth]{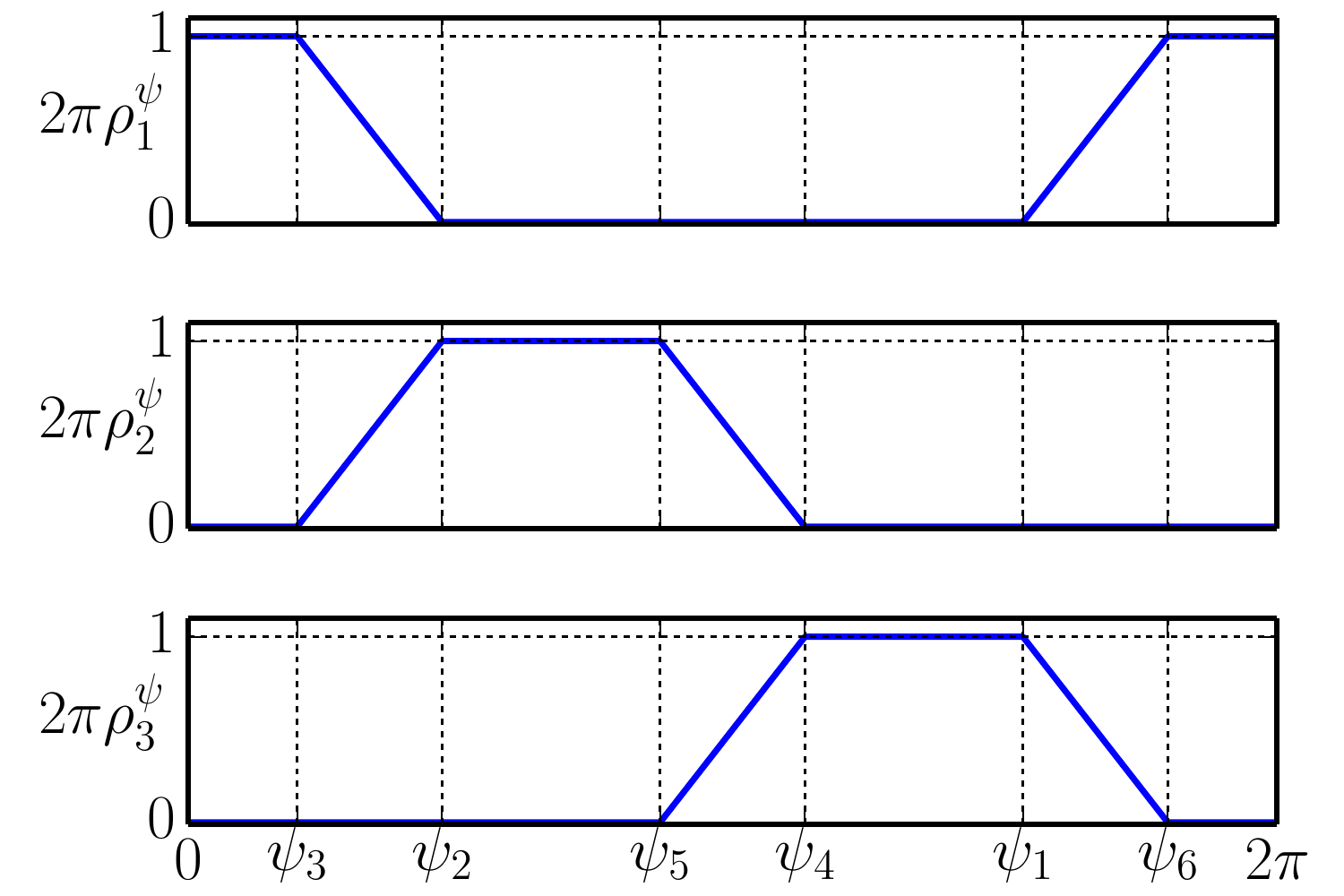}
\caption{ (Color online)
Density of particles in normal angle space as a function of the normal angle $\psi$  in each of the three convex lobes of the box shown in Fig.~\ref{fig:topology}.
$\psi_i$ is the normal angle of the inflexion point $A_i$ as well as that of its corresponding landing point $B_i$ (see Fig.~\ref{fig:topology}).
}
\label{fig:density_psi1}
\end{figure}

We start with the box pictured in Fig.~\ref{fig:topology}.
In order to write $\rhopsi(\psi_k)=(2\pi)^{-1}$ in terms of the unknowns $\{x_j\}$, we look at the normal angle representation (bottom right panel of Fig.~\ref{fig:topology}) and take a vertical slice at $\psi_k$.
Looking at, \eg, the slice through $B_5$, we see that $A_5$ and $B_5$ are the only two locations on the boundary where the normal angle is equal to $\psi_5$. At $B_5$, the density is by definition $x_5$, while at $A_5$ it is zero because it is the entrance of a concave region.
Therefore, the total density at $\psi_5$ is simply $x_5$, and the equation for that slice is $x_5=(2\pi)^{-1}$.
The situation is the same at every jump location, and the corresponding linear system is trivial:
\begin{align}
\forall k\in[1,2n], \quad x_k= \frac{1}{2\pi} 
\label{eq:linear_system1}
\end{align}
Recalling the piecewise linearity and boundary conditions $\rhopsi_i(\psi_{2i-1})=\rhopsi_i(\psi_{2i})=0$, the resulting partial densities $\rhopsi_i$ in $\psi$ space then take the form shown in Fig.~\ref{fig:density_psi1}.

\begin{figure} 
\centering
\includegraphics[width=\linewidth]{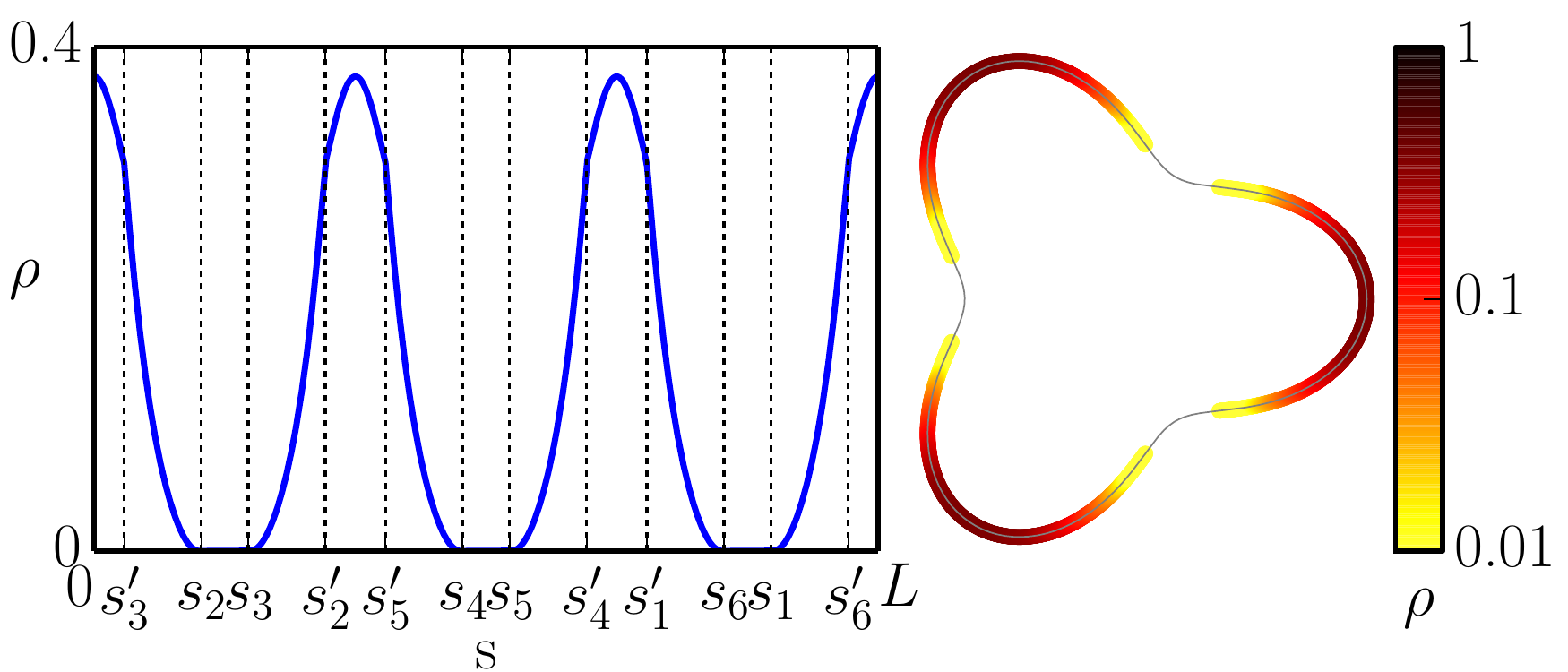}
\caption{ (Color online)
Density of particles per unit length of boundary as a function of arclength $s$ for the box shown in Fig.~\ref{fig:topology}.
$s_i$ and $s'_i$ are the arclengths of the inflexion points $A_i$ and corresponding landing points $B_i$ (see Fig.~\ref{fig:topology})).
Right panel: Heat map for the density along the boundary, shown in real space.
}
\label{fig:density1}
\end{figure}

Finally, Fig.~\ref{fig:density1} shows the density in real space and $s$ space, where the overlaps between convex regions disappear and the empty concave regions reappear.

More generally, let $m(\psi)$ be the multiplicity, \ie the number of convex locations that have normal angle $\psi$.
Graphically, in the normal angle representation (bottom right panel of Fig.~\ref{fig:topology}), $m(\psi)$ is the number of convex intervals that intersect the vertical line at $\psi$.
Since a jump connects two regions, $m(\psi_k) \ge 2$. If $m(\psi)$ never exceeds 2 over the entire boundary,
then $A_k$ and $B_k$ are the only convex locations with normal angle $\psi_k$, and the equation $\rhopsi(\psi_k)=(2\pi)^{-1}$ always takes the trivial form given by Eq.~\eqref{eq:linear_system1}, regardless of the number $n$ of convex regions or their sizes.

\subsubsection{An example with multiplicity $3$}
\label{multiplicity3}

\begin{figure} 
\centering
\includegraphics[width=0.95\linewidth]{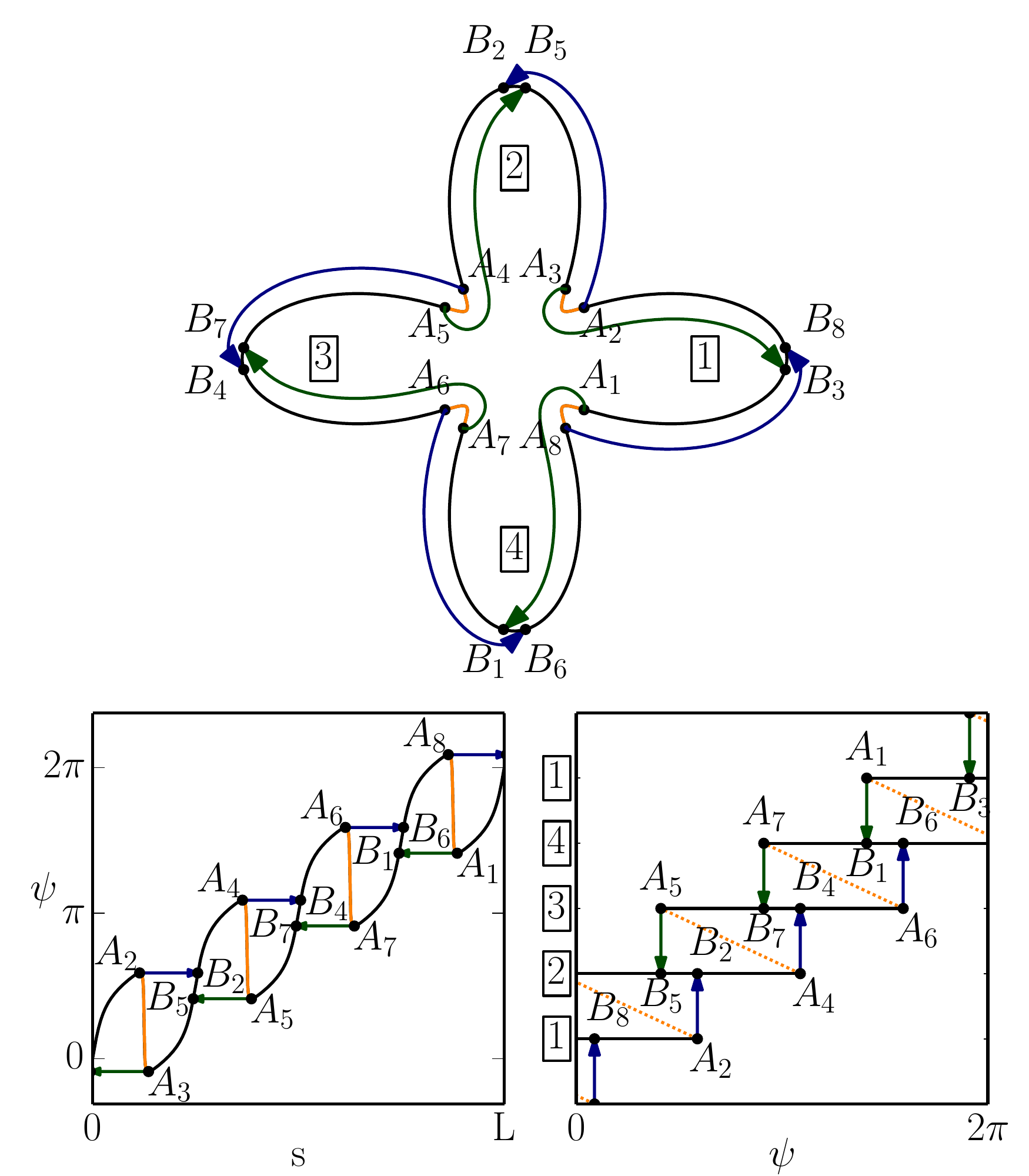}
\caption{ (Color online)
Three representations of a non-convex box showing the concave regions in orange (grey) and the  jumps over the concave regions (curved arrows). Convex regions are indexed by a number $1$ to $4$.
Top: Shape of the box.
The origin of arclengths is the rightmost point on the boundary, halfway between $B_3$ and $B_8$.
Bottom left: Normal angle vs. arclength.
Bottom right: Normal angle representation.
The vertical axis labels the convex region corresponding to each interval.
Region \region{1} appears split into two parts due to periodic boundary conditions.
}
\label{fig:topology2}
\end{figure}

We now consider the box shown in Fig.~\ref{fig:topology2}, defined in polar coordinates by
$r(\theta) = 1 + 0.6 \sin\left( 4\theta \right)$.

There are $n=4$ convex regions, delimited by the $8$ inflexion points $A_1$ to $A_8$.
The jump starting at $A_i$ ends at $B_i$, which is located in a neighboring lobe. Although the jumps  involve leaving the boundary and flying straight through the bulk for a short period of time (see section~\ref{ap:jump_typology}), these flights do not alter the location of the landing points $B_i$ and thus are irrelevant to the steady-state.

In real space, the most important difference between this box and the box of Fig.~\ref{fig:topology} is the order of the landing points. For example, $B_5$ comes before $B_2$ when moving counter-clockwise.
This ``inversion'' is a signature of  multiplicities higher than $2$, as can be seen by comparing the bottom panels of Figs.~\ref{fig:topology} and~\ref{fig:topology2}.

Following the method outlined in previous sections, we set out to write
$\rhopsi(\psi_k)=\sum_i\rhopsi_i(\psi_k)=(2\pi)^{-1}$ at each jump location $\psi_k$ in terms of the variables $x_k$.
We start with jump $7$ and draw a virtual vertical line through $A_7$ and $B_7$ in the normal angle representation (bottom right panel of Fig.~\ref{fig:topology2}).
This line intersects region $4$ at $A_7$, region $3$ at $B_7$, and region $2$ at a point $C$ located between $B_2$ and $A_4$.
We can therefore write
\begin{align}
(2\pi)^{-1}
 & = \rhopsi_4(\psi_7)+\rhopsi_3(\psi_7)+\rhopsi_2(\psi_7)
\nonumber \\
 & = x_7 + \frac{\psi_4-\psi_7}{\psi_4-\psi_2} \rhopsi_2(\psi_2)
           + \frac{\psi_7-\psi_2}{\psi_4-\psi_2} \rhopsi_2(\psi_4)
\nonumber \\
 & = x_7 + \frac{\psi_4-\psi_7}{\psi_4-\psi_2} x_2
\end{align}
The second line is obtained by using the boundary condition $\rhopsi_4(\psi_7)=0$, the definition $\rhopsi_3(\psi_7)=x_7$, and the linearity of $\rhopsi_2$ between $B_2$ and $A_4$.
The third line is obtained by using the boundary condition $\rhopsi_2(\psi_4)=0$ and the definition $\rhopsi_2(\psi_2)=x_2$.
Applying the method to each jump yields the system
\begin{align}
2\pi
\left(
\begin{array}{cccccccc}
1 & 0 & 0 & \alpha_1 & 0 & 0 & 0 & 0 \\
0 & 1 & 0 & 0 & \alpha_2 & 0 & 0 & 0 \\
0 & 0 & 1 & 0 & 0 & \alpha_3 & 0 & 0 \\
0 & 0 & 0 & 1 & 0 & 0 & \alpha_4 & 0 \\
0 & 0 & 0 & 0 & 1 & 0 & 0 & \alpha_5 \\
\alpha_6 & 0 & 0 & 0 & 0 & 1 & 0 & 0 \\
0 & \alpha_7 & 0 & 0 & 0 & 0 & 1 & 0 \\
0 & 0 & \alpha_8 & 0 & 0 & 0 & 0 & 1
\end{array}
\right)
\cdot
\left(
\begin{array}{c}
x_1 \\ x_2 \\ x_3 \\ x_4 \\ x_5 \\ x_6 \\ x_7 \\ x_8
\end{array}
\right)
=
\left(
\begin{array}{c}
1 \\ 1 \\ 1 \\ 1 \\ 1 \\ 1 \\ 1 \\ 1
\end{array}
\right)
\end{align}
where
$\alpha_k = \dfrac{\psi_{k-3}-\psi_k}{\psi_{k-3}-\psi_{k-5}}$
(indices are defined modulo $2n$).
The solution to this system is not in general compact; however, here the four-fold symmetry implies that $\alpha_k\equiv\alpha$ is independent of $k$,and the solution simply reads:
\begin{align}
\forall k , \  x_k = \frac{1}{2\pi (1+\alpha)}
\label{eq:multiplicity3_result}
\end{align}

\begin{figure} 
\centering
\includegraphics[width=0.8\linewidth]{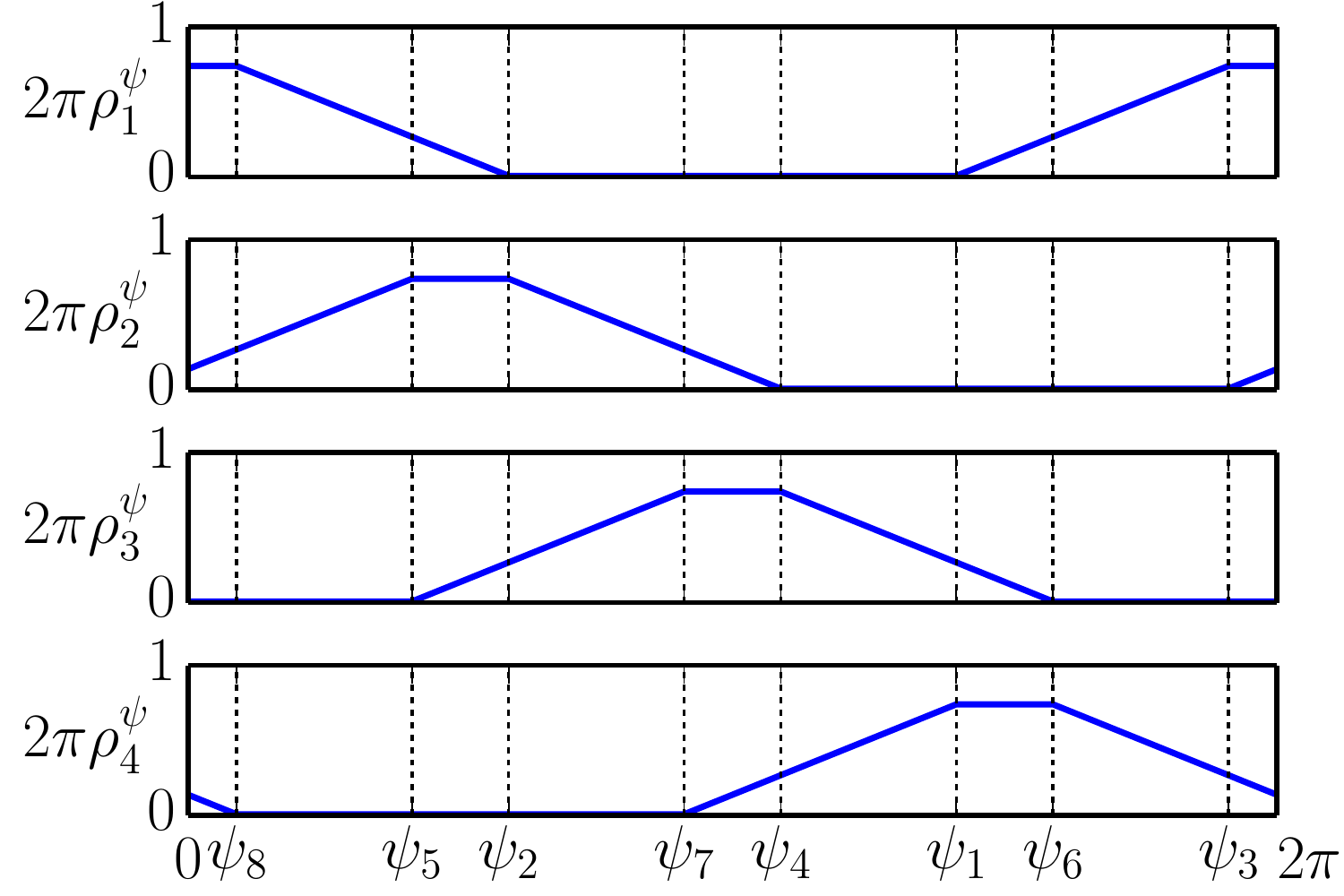}
\caption{ (Color online)
Density of particles in normal angle space as a function of the normal angle $\psi$ in each of the four convex lobes of the box shown in Fig.~\ref{fig:topology2}.
}
\label{fig:density_psi2}
\end{figure}

\begin{figure} 
\centering
\includegraphics[width=\linewidth]{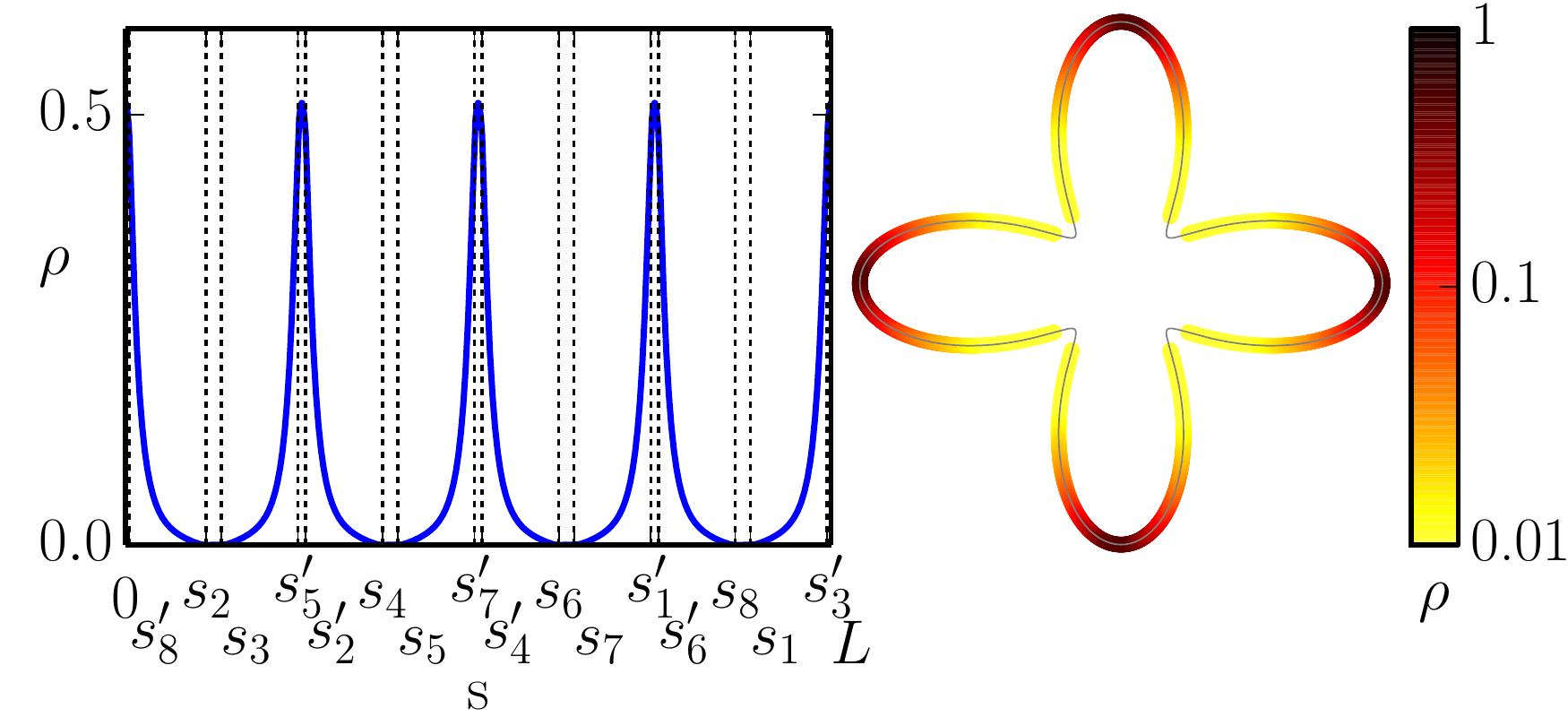}
\caption{ (Color online)
Density of particles per unit length of boundary as a function of arclength $s$ for the box shown in Fig.~\ref{fig:topology2}.
Right panel:
Heat map for the density along the boundary, shown in real space.
}
\label{fig:density2}
\end{figure}

The corresponding densities in normal angle space, in arclength space and in real space are shown in Figs.~\ref{fig:density_psi2} and~\ref{fig:density2}.

\subsubsection{General case}
\label{general_case}

The method used above to compute the steady-state density in simple boxes can be readily extended to arbitrary shapes. We now write down the steps leading to the solution in the general case.

As a preliminary step, in each region $i$
we identify the set $K_i = \{ k \in [1,2n]\ |\ s'_k \in [s_{2i-1},s_{2i}] \}$ of jumps that land in the region, with $s'_k$ the arclength of landing point $B_k$.
We also define the set ${\cal E}_i=\{A_{2i-1},A_{2i}\}\cup\{B_k\ |\ k \in K_i\}$ of jump ends in region $i$, \ie the two ends of the region plus any jump landings.
Then, for each jump landing $B_k$, we perform the following sequence of operations: \\
\begin{enumerate}
\setlength{\itemsep}{0pt}
\setlength{\parsep}{0pt}
\item Identify the region $i$ that contains $B_k$. By definition $\rhopsi_i(\psi_k)=x_k$. \\
\item Pick a region $j$ other than $i$. In the set ${\cal E}_j$, identify the two points $C$ and $D$ whose normal angles are closest to $\psi_k$ on each side:
\begin{align*}
\psi_C = \max_{M\in{\cal E}_j} \{ \psi_M\ |\ \psi_M\le\psi_k \} \\
\psi_D = \min_{M\in{\cal E}_j} \{ \psi_M\ |\ \psi_M\ge\psi_k \}
\end{align*}
By construction, $\rhopsi_j$ is linear between $C$ and $D$; therefore
\begin{align}
\rhopsi_j(\psi_k) = \frac{\psi_D-\psi_k}{\psi_D-\psi_C} \rhopsi_j(\psi_C)
               + \frac{\psi_k-\psi_C}{\psi_D-\psi_C} \rhopsi_j(\psi_D)
\label{eq:build_solution1}.
\end{align}
Furthermore, $\rhopsi_j(\psi_{C/D})$ is either $0$ if $C/D$ is $A_{2j-1}$ or $A_{2j}$, or $x_l$ if $C/D$ is $B_l$; therefore the right-hand side of Eq.~\eqref{eq:build_solution1} is a linear combination of $0$, $1$ or $2$ of the $x_l$'s (if the region has no jump landing, or if $\psi_k$ is outside of the region, the right-hand side is simply zero). \\
\item Repeat the previous step until every region has been considered, then sum Eq.~\eqref{eq:build_solution1} over $j$. The left-hand side is (by definition) $\rhopsi(\psi_k)=(2\pi)^{-1}$, while the right-hand side is a linear combination of $x_l$'s.
\end{enumerate}
At the end of step 3, a linear system of the form $\sum_j a_{ij} x_j = (2\pi)^{-1}$ is obtained, whose coefficients $a_{ij}$ depend on the $\psi_k$'s.
After inverting the system to get the $x_j$'s and thus the normal angle space densities $\{\rhopsi_i(\psi)\}$, ``unfolding'' normal angle space onto arclength space and dividing by the radius of curvature yields the density $\rho(s)$ per unit length of boundary (see Eq.~\eqref{eq:unfold}).

As pointed out at the beginning of section~\ref{quasi_density}, the process can be automated. This is particularly useful when the number of non-zero elements in $a_{ij}$ is large, which happens when the multiplicity $m$ is large.
On the other hand, a small multiplicity leads to a sparse matrix. In particular,  $m=2$ yields a diagonal matrix (see section~\ref{multiplicity2}).

\section{Simulations}
\label{simulations}

We test the results of sections~\ref{wall_dynamics} and~\ref{quasi_density} by performing molecular dynamics simulations of Eqs.~\eqref{eq:eom1} in the family of boxes defined in polar coordinates by
\begin{align}
r(\theta) = 1 + r_1 \sin\left( k \theta \right)
\label{eq:polar_eq}
\end{align}
The boxes shown in Figs.~\ref{fig:topology} and~\ref{fig:topology2} belong to this family, with $(k,r_1)=(2,0.5)$ and $(4,0.6)$ respectively.

\subsection{Angular distribution}

We first test our core assumption, that the deviation of the particle orientation from the boundary normal vanishes: $\phi\approx0$, and the reasoning that led to it (see section~\ref{wall_dynamics}), by characterizing the statistics of $\phi$ at various locations along the boundary.
In convex regions, the theory predicts that the distribution of $\phi$ will exhibit a narrow peak centered around $\phi=0$. 
On the other hand, particles jumping over a concave region should generate secondary peaks, much smaller than the primary peak, and centered around a position-dependent non-zero value $\phi=\psi_0-\psi$ where $\psi_0$ and $\psi$ are the normal angles at the inflexion point where the particle entered the jump and at the current location, respectively.
In concave regions, there should be no central peak, only ``secondary'' ones.

\begin{figure}[h]
\centering
\includegraphics[width=\linewidth]{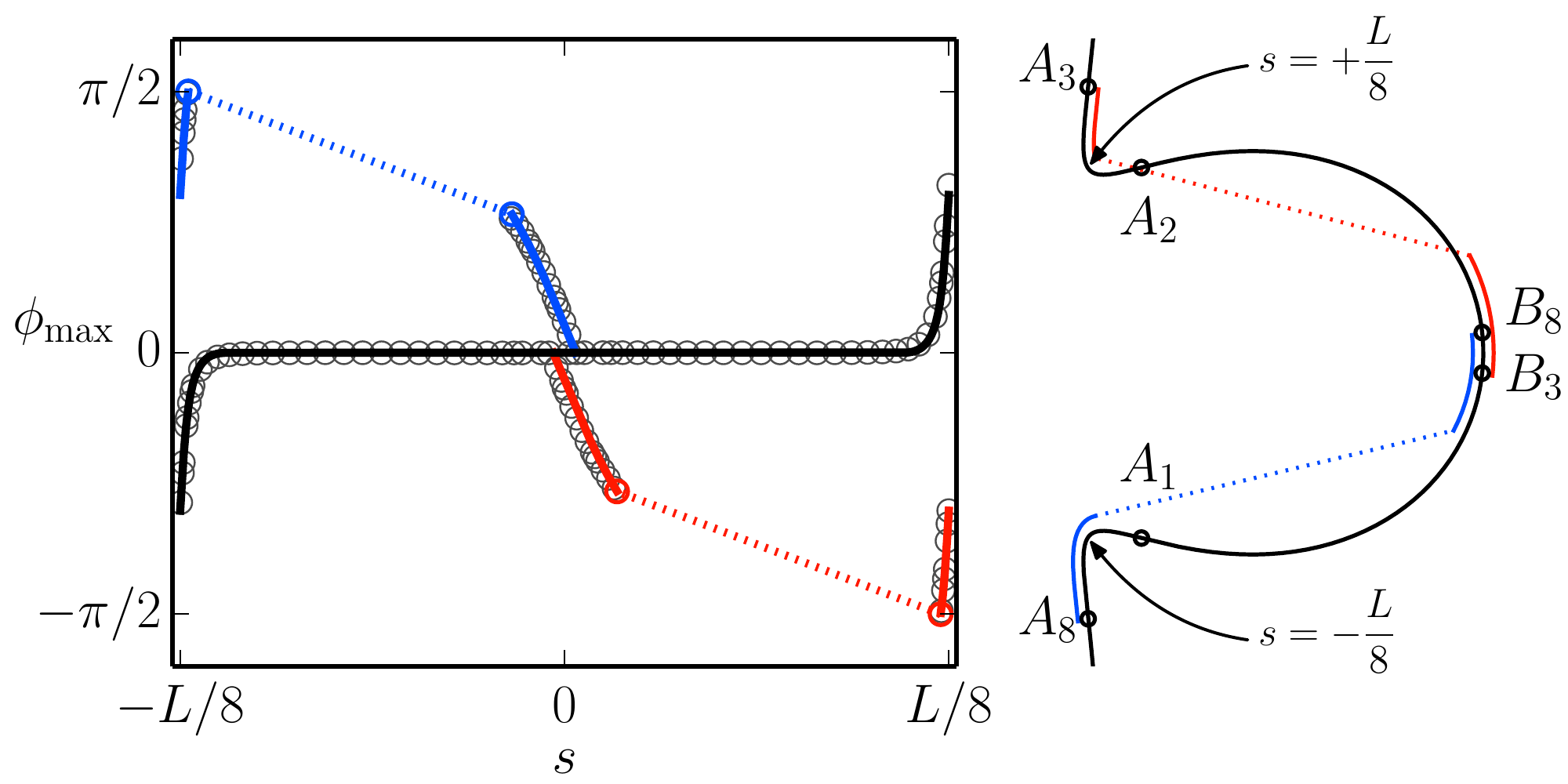}
\caption{ (Color online)
Positions of peaks in the distribution of the orientations $P(\phi)$ relative to the normal as a function of the arclength $s$ over one lobe of the box defined by Eq.~\eqref{eq:polar_eq} with $k=4$ and $r_1=0.5$. The lobe geometry is shown in the right panel.
The angular diffusion constant is $\Dr=10^{-4}$, deep in the strong confinement regime.
The simulation data is shown as circles.
The theoretical prediction is shown as lines. Colors denote the origin of the particles: blue for particles jumping from $A_8$ (upper gray curve), orange for particles jumping from $A_3$ (lower gray curve), and black for particles from the lobe under consideration.
Dotted lines represent the paths of particles that fly through the bulk.
}
\label{fig:orientation_peaks}
\end{figure}

To compare these predictions with the simulation results, we measure the distribution $P(\phi)$ at regularly spaced locations along the boundary. At each point, we determine the heights of the distribution's peaks $\{P_\text{max}\}$ and their corresponding orientations $\{\phi_\text{max}\}$. In Fig.~\ref{fig:orientation_peaks} we show the peak orientations as a function of arclength in a single convex region, as well as the prediction associated with each jump involving the region.
Every detected peak falls on one of the predicted branches: $\phi=0$ or $\phi=\psi_i-\psi$ with $i\in{1,2,3,8}$ depending on whether the corresponding jump started at $A_1$, $A_2$, $A_3$, or $A_8$.
Fig.~\ref{fig:orientation_peaks} also illustrates particles leaving the boundary to fly straight through the bulk (see middle row of Fig.~\ref{fig:concave_jumps}):
since $\phi$ is only defined at the boundary, the peak corresponding to these particles is absent between their leaving the boundary and their hitting it again (dotted lines in Fig.~\ref{fig:orientation_peaks}).

\subsection{Steady-state density}

\newcommand{\rhopr}{\rho_\text{pr}}

\begin{figure}[h]
\centering
\includegraphics[width=\linewidth]{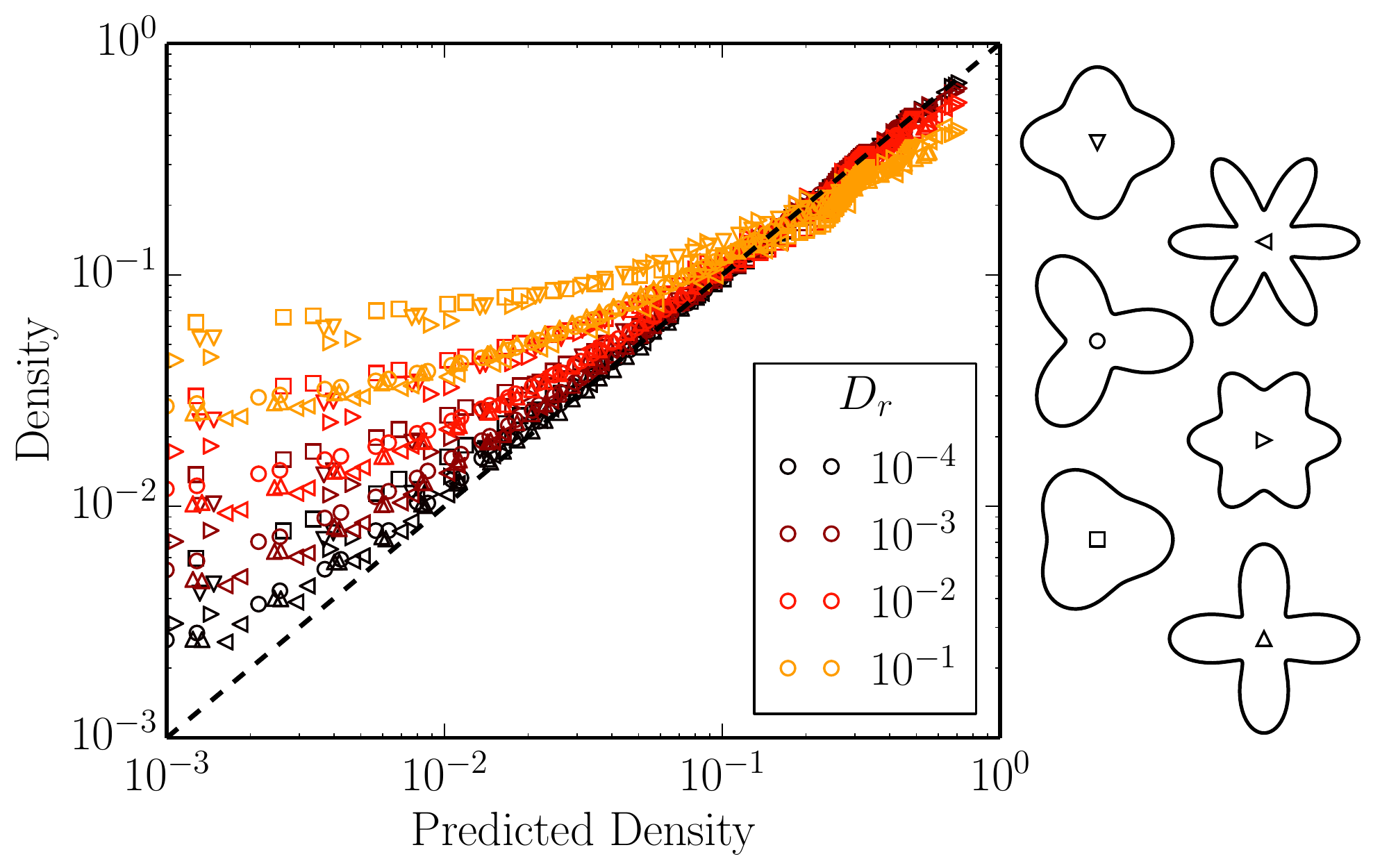}
\caption{ (Color online)
The density observed in simulations at various positions along the boundary is plotted against the predicted density for six boxes from the family defined by Eq.~\eqref{eq:polar_eq} and for four values of the angular diffusion constant $\Dr$.
The dashed line corresponds to a density equal to the predicted density.
The symbol associated with each box geometry is shown at the center of that box on the right side.
}
\label{fig:density_vs_prediction}
\end{figure}

We now assess the accuracy of the predictions made in section~\ref{quasi_density} by plotting in Fig.~\ref{fig:density_vs_prediction} the observed boundary density $\rho$ as a function of the predicted density $\rhopr$ for various box shapes and angular diffusion constants $\Dr$.

The plot can be interpreted in terms of two linear asymptotes.
 In boundary locations with moderate and high density, we observe good agreement between theory and simulation, \ie $\rho\approx\rhopr$, up to $\Dr\sim10^{-2}$.
Deep in the strong confinement regime ($\Dr=10^{-4}$), the agreement is excellent and persists over two decades. 
In locations with low density, on the other hand, the prediction underestimates the density:
the predicted density vanishes altogether over regions of zero or negative curvature while the observed density is finite everywhere,
resulting in a horizontal asymptote whose position depends on the box's shape and the angular noise's strength.
Note that because of the logarithmic scale, regions where the predicted density is zero do not appear in Fig.~\ref{fig:density_vs_prediction}; instead the data visible on the left of the plot comes from weakly convex areas near the inflexion points.

As discussed in section~\ref{quasi_dynamics} and appendix~\ref{ap:jump_duration}, the density in flat and concave regions is controlled by the ratio of the time spent crossing them to the reorientation time $\Dr^{-1}$.
This ratio only vanishes in the $\Dr\rightarrow0$ limit,
therefore a finite density is to be expected in those regions at finite $\Dr$.
Since the crossing time is largest in flat regions, this is where the deviations from the quasi-static theory are most prominent.
Additionally, the time it takes to cross the vicinity of an inflexion point grows with its ``flatness'', as inferred from the second derivative of the normal angle with respect to the arclength (\ie the derivative of the curvature).
This explains the potentially counter-intuitive observation that the accuracy of our predicted density is better in ``strongly concave'' boxes [\eg the box denoted by left-pointing triangles ($\triangleleft$) in Fig.~\ref{fig:density_vs_prediction}] than in ``weakly concave'' boxes [\eg the box denoted by squares ($\square$)]. Despite their weaker concavity, the latter exhibit larger ``flat'' regions.

\begin{figure}[h]
\centering
\includegraphics[width=\linewidth]{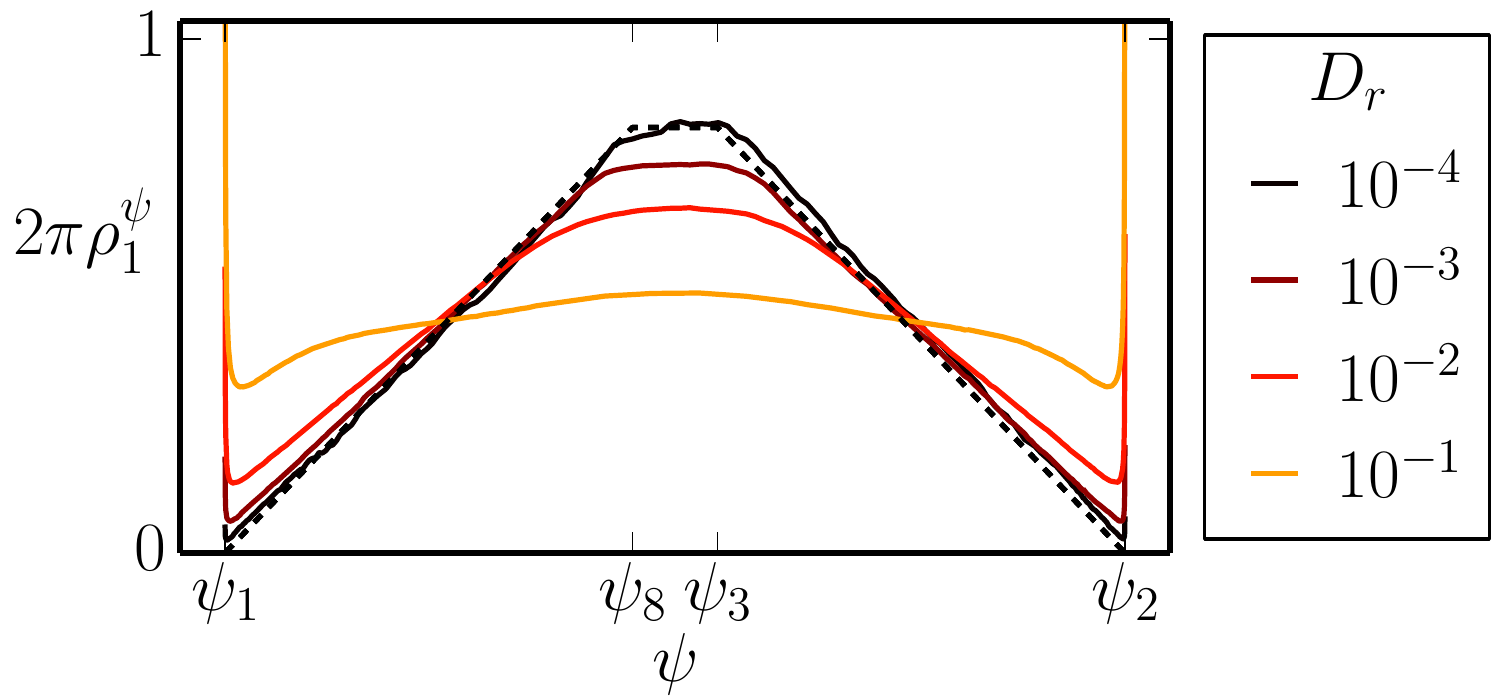}
\caption{ (Color online)
Density in normal angle space in the convex region shown in Fig.~\ref{fig:orientation_peaks} for several values of the angular diffusion constant $\Dr$ from deep in the strong confinement regime ($\Dr=10^{-4}$) to outside of it ($\Dr=10^{-1}$).
The dashed line is the theoretical prediction from section~\ref{multiplicity3} (see Eq.~\eqref{eq:multiplicity3_result}), corresponding to $\Dr\rightarrow0$.
}
\label{fig:density_psi_sim}
\end{figure}

Finally, we plot in Fig.~\ref{fig:density_psi_sim} the density $\rhopsi=R\rho$ in normal angle ($\psi$) space.
There, each convex region must be treated separately; we consider the region labelled \framebox{1} in Fig.~\ref{fig:topology2}.
The predicted density is piecewise linear with a trapezoidal shape.
At very small noise ($\Dr=10^{-4}$), the density observed in simulations closely matches the prediction.
As the noise is increased, the trapezoidal shape gets smoothed out and some of the density is transferred from the tip to the base (as well as the neighboring concave region, not shown in this representation),
suggesting that finite noise effects may be treated as a perturbation to our zero-noise theory.
On the other hand, at $\Dr=10^{-1}$ the predicted form of the density is not recognizable anymore and a different approach is required.
Note that this description over-emphasizes the finiteness of the observed density at the ends of the convex interval where the radius of curvature, and thus $\rhopsi$, diverges.

\section{Discussion}
\label{discussion}

In summary, we have presented a systematic approach to predict the density of a non-aligning ideal active gas in a small box of arbitrary shape, thus establishing a connection between the geometry of a confining box and the properties of the active gas it confines.
Our results hold as long as the persistence length (the distance a free particle travels before it loses its orientation) is much larger than the size of the box.

In the special case of convex boxes, there is a strikingly simple relationship between the density and the boundary geometry: the density is zero in the bulk and on the boundary it is proportional to the local boundary curvature~\cite{Fily2014a}.
Here, we have shown that boundaries with concave regions lead to a much richer particle dynamics, including multi-stability, hysteretic dynamics, and particles flying through the bulk of the box between disparate boundary locations. However, we showed that the particle density still vanishes in the bulk and we described an algorithm to calculate the steady-state density profile on the boundary of a 2D box with any shape. The predicted particle density vanishes in concave regions, while in convex regions it can be written as the product of the local curvature and a ``splitting factor'' which obeys the following property:
given a unit vector $\vecu{n}$, the sum of the splitting factor over all the locations on the boundary where the normal is $\vecu{n}$ is equal to one.
In other words, boundary points that share the same normal also share the same ``pool'' of particles with the corresponding orientation.

Despite the complexity intrinsic to concave regions, understanding non-convex shapes is essential to rationally design active micro-devices with specific functionalities.
This is nicely illustrated by the micro-gears used in Refs.~\cite{Angelani2009,DiLeonardo2010,Sokolov2010},
which trap particles in sharp corners where they exert torques that make the gear rotate~%
\footnote{%
Note that the theory developed in this paper only describes particles trapped inside a box, whereas Refs.~\cite{Angelani2009,DiLeonardo2010} are concerned with active particles swimming outside the gear.
However, Ref.~\cite{Sokolov2010} demonstrates that both configurations lead to a torque on the gear using very similar shapes, therefore the question of how to design the gear may be discussed from either point of view.
}.
Effectively trapping active particles requires sharp corners~\cite{Kaiser2012,Kaiser2013,Guidobaldi2014,Fily2014a}, and the total torque on the gear is maximized by having several such trapping sites.
This cannot, however, be achieved with convex shapes in which the number of sharp corners is limited to two.
It is then clear that understanding and using non-convex confinements is a necessary step toward designing a broad class of  active devices.

{\bf Scope of the model.} Finally, we consider limitations and avenues for extension of the model. Firstly, since we neglect interparticle interactions our results are limited to dilute systems. For example, above a threshold packing fraction, steric effects will prevent all particles from residing on the boundary.  The effect of steric interactions will be discussed in a future publication, but preliminary simulations confirm that our results apply at least qualitatively at finite particle densities.
Secondly, our results apply to the strong confinement limit, in which particles circumnavigate the box faster than they reorient and thus tend to align with the boundary normal.  Within this limit, we expect the results to remain valid regardless of the reorientation mechanism, including angular diffusion or aligning interactions with the wall such as may arise due to hydrodynamics. However, if particle-wall interactions drive particles to align with the wall, or divert away from it, on timescales comparable to the circumnavigation time ($\sim \vel_0/R$ with $R$ the boxsize), then a different approach is required.

\appendix

\section{Types of jumps}
\label{ap:jump_typology}

\begin{figure}
\centering
\includegraphics[width=0.8\linewidth]{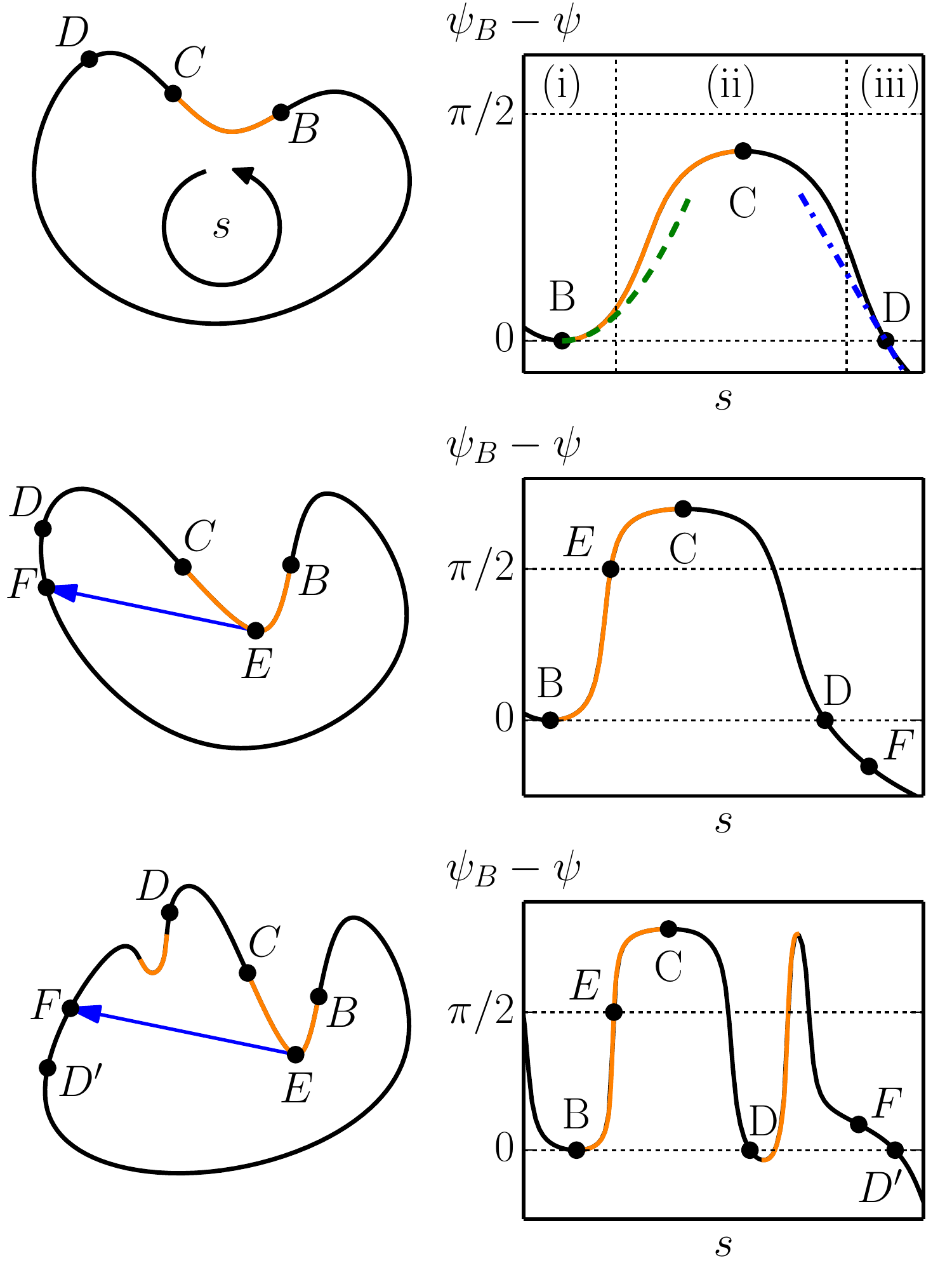}
\caption{ (Color online)
Types of jumps over a concave region.
The right column shows the orientation $\phi=\psi_B-\psi$ relative to the boundary as a function of the arclength $s$ (counted counter-clockwise).
Concave regions are in orange (grey).
Trajectories through the interior of the box are shown as arrows.
$B$ and $C$ are the inflexion points delimiting the concave region.
$D$ is the first location after $C$ with the same normal as $B$.
Top row: the particle follows the boundary and stops at $D$.
See section~\ref{ap:jump_duration} for the definition of regions (i), (ii), (iii).
The dashed (in green) and the dot-dashed (in blue) curves correspond to a quadratic expansion near $B$ and a linear expansion near $D$, respectively.
Middle row: when $\psi_B-\psi_C>\pi/2$, the particle leaves the boundary at $E$ where $\psi_E=\psi_B-\pi/2$. It travels in a straight line through the bulk until $F$, then follows the boundary to $D$.
Bottom row: when $\psi_B-\psi_C>\pi/2$ and there are multiple concave regions, the particle may fly past $D$ to end its jump in a non-neighboring convex region at a point $D'$  with the same normal as $B$.
}
\label{fig:concave_jumps}
\end{figure}

Depending on the geometry of the concave region they cross, jumps may lead to particles leaving the boundary to travel in the interior of the box.
This situation, which is illustrated in Fig.~\ref{fig:concave_jumps}, occurs when the angle between the normals at the two ends of the concave region ($B$ and $C$) is larger than $\pi/2$.
As a result, there exists a point $E$ between $B$ and $C$ where $\phi=\psi_B-\psi_E=\pi/2$. At $E$, the particle's orientation is aligned with the tangent and it leaves the boundary to travel in a near-straight line through the box, until it hits the boundary again.
The three possible jump scenarios are explained in Fig.~\ref{fig:concave_jumps}.

\section{Jump duration}
\label{ap:jump_duration}

To evaluate the duration of a jump, we consider the box shown in the top panel of Fig.~\ref{fig:concave_jumps} and a particle at the entrance $B$ of the concave region with orientation $\theta=\psi_B+d\theta$ where $d\theta>0$ is an infinitesimal forward velocity needed to start the jump.
In the absence of angular noise ($\Dr=0$), the gliding speed $\dot{s}=v_0\sin\phi$ is controlled by the relative angle $\phi=\theta-\psi\approx\psi_B-\psi$, shown in the top right panel of Fig.~\ref{fig:concave_jumps}.

In particular, we want to show that in the small angular noise limit $\Dr\rightarrow0$, the duration of the jump from $B$ to $D$ is negligible compared with the reorientation time $\tau_0\equiv\Dr^{-1}$.

To this end, we divide the jump into three regions numbered (i), (ii) and (iii) (see top right panel of Fig.~\ref{fig:concave_jumps}), and define $\tau_\alpha$ as the time it takes to cross region $\alpha$.
Regions (i) and (iii) correspond to the vicinity of $B$ and $D$, respectively, while region (ii) contains the remaining middle part of the jump.

Region (ii) is the fastest of the three, with a relative angle $\phi\sim1$ and a gliding speed $\dot{s}\sim v_0$. This leads to a crossing time $\tau_{ii}\sim \ell/v_0$ where $\ell$ is the length of the jump, which is of the same order or smaller as the size of the box.
Note that instances of the particle leaving the boundary (see section~\ref{ap:jump_typology}, and the middle and bottom row of Fig.~\ref{fig:concave_jumps}) do not modify the scaling of $\tau_{ii}$. Indeed, when flying through the box the particle travels at speed $v_0$ for a distance at most of  the order of the box size.
Comparing with the reorientation time, we get $\tau_{ii}/\tau_0\sim \ell\Dr/v_0$, which vanishes in the small noise limit.

In region (iii), linearizing the gliding velocity around $D$ leads to Eq.~\eqref{eq:eom5} with $s_0=s_D$ the arclength of point $D$.
The dynamics is an exponential relaxation towards $D$, which takes a time $\tau_{iii}\sim R_D/v_0$ where $R_D$ is the radius of curvature at point $D$~\footnote{Technically, $D$ is never reached. However, the particle gets within any reasonably small distance of $D$ within a few relaxation times.}.
Comparing with the reorientation time, we get $\tau_{iii}/\tau_0\sim R_D\Dr/v_0$, which also vanishes in the small noise limit.

Region (i) is the slowest part of the jump. While $B$ is a fixed point, it is also an inflexion point, and thus the linear contribution in Eq.~\eqref{eq:eom5} vanishes. A second order expansion yields
$\dot{x} = v_0 d\theta + v_0\kappa x^2$
where $x=s-s_B$, $\kappa=\frac12 (d^2\psi/ds^2)$, and $v_0 d\theta$ is the initial velocity required to start the jump.
Integrating leads to the displacement
%
\begin{align}
x(t)=\sqrt{\frac{d\theta}{\kappa}}\tan\left(v_0 t\sqrt{\kappa d\theta}\right)
\end{align}
However, evaluating $\tau_i$ requires additional assumptions on $d\theta$. In fact, noise and curvature both play a role in setting $\tau_i$. Although noise is initially the only contribution, its fluctuations soon get amplified by curvature effects, and understanding their interplay is necessary to evaluate $\tau_i$.

For our purpose, however, it is sufficient to note that the curvature in region (i), however small, always makes the particle progress faster than on a flat edge. The case of a flat edge was discussed in Ref.~\cite{Fily2014a} in the context of polygonal boxes, and we briefly summarize it here.
In the absence of curvature, the relative angle $\phi$ follows a random walk. Its typical value then grows diffusively: $\phi\sim(\Dr t)^{1/2}$.
Integrating with respect to time gives the typical distance travelled along the boundary after a time $t$: $s\sim v_0 \Dr^{1/2} t^{3/2}$.
Travelling a length $\ell$ then takes a typical time $\tau\sim(\ell/v_0)^{2/3} \Dr^{-1/3}$.
The length of region (i) can be evaluated by noting it ends when $\psi-\psi_B\approx\kappa x^2$ is of order $1$.
This gives $\ell\approx\kappa^{-1/2}$ and
$\tau_{i}\sim (v_0^2\kappa\Dr)^{-1/3}$.
Comparing with the reorientation time, we get
$\tau_i/\tau_0\sim (v_0 \Dr)^{2/3} (\kappa)^{-1/3}$,
which goes to zero in the small noise limit as well.

In summary, in the small noise limit $\Dr\rightarrow0$ jumps over concave regions happen much faster than the particles reorient.
This in turn ensures the consistency of the quasi-static approach to steady-state density in non-convex boxes.
Additionally, the duration of the jump is controlled by its slowest and earliest stage, near the inflexion point, which grows with the inverse of the second derivative $(d^2\psi/ds^2)$ of the normal angle with respect to arclength.
In other words, the flatter the region near inflexion points, the slower the associated jumps.

\begin{acknowledgments}
This research was supported by
NSF-MRSEC-0820492 and NSF-DMR-1149266.
Computational resources were provided by the NSF through XSEDE computing resources and the
Brandeis HPCC.
\end{acknowledgments}

\bibliography{confined_active2}

\begin{thebibliography}{51}%
\makeatletter
\providecommand \@ifxundefined [1]{%
 \@ifx{#1\undefined}
}%
\providecommand \@ifnum [1]{%
 \ifnum #1\expandafter \@firstoftwo
 \else \expandafter \@secondoftwo
 \fi
}%
\providecommand \@ifx [1]{%
 \ifx #1\expandafter \@firstoftwo
 \else \expandafter \@secondoftwo
 \fi
}%
\providecommand \natexlab [1]{#1}%
\providecommand \enquote  [1]{``#1''}%
\providecommand \bibnamefont  [1]{#1}%
\providecommand \bibfnamefont [1]{#1}%
\providecommand \citenamefont [1]{#1}%
\providecommand \href@noop [0]{\@secondoftwo}%
\providecommand \href [0]{\begingroup \@sanitize@url \@href}%
\providecommand \@href[1]{\@@startlink{#1}\@@href}%
\providecommand \@@href[1]{\endgroup#1\@@endlink}%
\providecommand \@sanitize@url [0]{\catcode `\\12\catcode `\$12\catcode
  `\&12\catcode `\#12\catcode `\^12\catcode `\_12\catcode `\%12\relax}%
\providecommand \@@startlink[1]{}%
\providecommand \@@endlink[0]{}%
\providecommand \url  [0]{\begingroup\@sanitize@url \@url }%
\providecommand \@url [1]{\endgroup\@href {#1}{\urlprefix }}%
\providecommand \urlprefix  [0]{URL }%
\providecommand \Eprint [0]{\href }%
\providecommand \doibase [0]{http://dx.doi.org/}%
\providecommand \selectlanguage [0]{\@gobble}%
\providecommand \bibinfo  [0]{\@secondoftwo}%
\providecommand \bibfield  [0]{\@secondoftwo}%
\providecommand \translation [1]{[#1]}%
\providecommand \BibitemOpen [0]{}%
\providecommand \bibitemStop [0]{}%
\providecommand \bibitemNoStop [0]{.\EOS\space}%
\providecommand \EOS [0]{\spacefactor3000\relax}%
\providecommand \BibitemShut  [1]{\csname bibitem#1\endcsname}%
\let\auto@bib@innerbib\@empty
\bibitem [{\citenamefont {Dombrowski}\ \emph {et~al.}(2004)\citenamefont
  {Dombrowski}, \citenamefont {Cisneros}, \citenamefont {Chatkaew},
  \citenamefont {Goldstein},\ and\ \citenamefont {Kessler}}]{Dombrowski2004}%
  \BibitemOpen
  \bibfield  {author} {\bibinfo {author} {\bibfnamefont {C.}~\bibnamefont
  {Dombrowski}}, \bibinfo {author} {\bibfnamefont {L.}~\bibnamefont
  {Cisneros}}, \bibinfo {author} {\bibfnamefont {S.}~\bibnamefont {Chatkaew}},
  \bibinfo {author} {\bibfnamefont {R.~E.}\ \bibnamefont {Goldstein}}, \ and\
  \bibinfo {author} {\bibfnamefont {J.~O.}\ \bibnamefont {Kessler}},\
  }\href@noop {} {\bibfield  {journal} {\bibinfo  {journal} {Phys. Rev. Lett.}\
  }\textbf {\bibinfo {volume} {93}},\ \bibinfo {pages} {098103 (4 pages)}
  (\bibinfo {year} {2004})}\BibitemShut {NoStop}%
\bibitem [{\citenamefont {Peruani}\ \emph {et~al.}(2012)\citenamefont
  {Peruani}, \citenamefont {Starru{\ss}}, \citenamefont {Jakovljevic},
  \citenamefont {S{\o}gaard-Andersen}, \citenamefont {Deutsch},\ and\
  \citenamefont {B\"ar}}]{Peruani2012}%
  \BibitemOpen
  \bibfield  {author} {\bibinfo {author} {\bibfnamefont {F.}~\bibnamefont
  {Peruani}}, \bibinfo {author} {\bibfnamefont {J.}~\bibnamefont
  {Starru{\ss}}}, \bibinfo {author} {\bibfnamefont {V.}~\bibnamefont
  {Jakovljevic}}, \bibinfo {author} {\bibfnamefont {L.}~\bibnamefont
  {S{\o}gaard-Andersen}}, \bibinfo {author} {\bibfnamefont {A.}~\bibnamefont
  {Deutsch}}, \ and\ \bibinfo {author} {\bibfnamefont {M.}~\bibnamefont
  {B\"ar}},\ }\href {\doibase 10.1103/PhysRevLett.108.098102} {\bibfield
  {journal} {\bibinfo  {journal} {Phys. Rev. Lett.}\ }\textbf {\bibinfo
  {volume} {108}},\ \bibinfo {pages} {098102} (\bibinfo {year}
  {2012})}\BibitemShut {NoStop}%
\bibitem [{\citenamefont {Poujade}\ \emph {et~al.}(2007)\citenamefont
  {Poujade}, \citenamefont {Grasland-Mongrain}, \citenamefont {Hertzog},
  \citenamefont {Jouanneau}, \citenamefont {Chavrier}, \citenamefont {Ladoux},
  \citenamefont {Buguin},\ and\ \citenamefont {Silberzan}}]{Poujade2007}%
  \BibitemOpen
  \bibfield  {author} {\bibinfo {author} {\bibfnamefont {M.}~\bibnamefont
  {Poujade}}, \bibinfo {author} {\bibfnamefont {E.}~\bibnamefont
  {Grasland-Mongrain}}, \bibinfo {author} {\bibfnamefont {A.}~\bibnamefont
  {Hertzog}}, \bibinfo {author} {\bibfnamefont {J.}~\bibnamefont {Jouanneau}},
  \bibinfo {author} {\bibfnamefont {P.}~\bibnamefont {Chavrier}}, \bibinfo
  {author} {\bibfnamefont {B.}~\bibnamefont {Ladoux}}, \bibinfo {author}
  {\bibfnamefont {A.}~\bibnamefont {Buguin}}, \ and\ \bibinfo {author}
  {\bibfnamefont {P.}~\bibnamefont {Silberzan}},\ }\href@noop {} {\bibfield
  {journal} {\bibinfo  {journal} {PNAS}\ }\textbf {\bibinfo {volume} {104}},\
  \bibinfo {pages} {15988} (\bibinfo {year} {2007})}\BibitemShut {NoStop}%
\bibitem [{\citenamefont {Trepat}\ \emph {et~al.}(2009)\citenamefont {Trepat},
  \citenamefont {Wasserman}, \citenamefont {Angelini}, \citenamefont {Millet},
  \citenamefont {Weitz}, \citenamefont {Butler},\ and\ \citenamefont
  {Fredberg}}]{Trepat2009}%
  \BibitemOpen
  \bibfield  {author} {\bibinfo {author} {\bibfnamefont {X.}~\bibnamefont
  {Trepat}}, \bibinfo {author} {\bibfnamefont {M.~R.}\ \bibnamefont
  {Wasserman}}, \bibinfo {author} {\bibfnamefont {T.~E.}\ \bibnamefont
  {Angelini}}, \bibinfo {author} {\bibfnamefont {E.}~\bibnamefont {Millet}},
  \bibinfo {author} {\bibfnamefont {D.~A.}\ \bibnamefont {Weitz}}, \bibinfo
  {author} {\bibfnamefont {J.~P.}\ \bibnamefont {Butler}}, \ and\ \bibinfo
  {author} {\bibfnamefont {J.~J.}\ \bibnamefont {Fredberg}},\ }\href@noop {}
  {\bibfield  {journal} {\bibinfo  {journal} {Nat. Phys.}\ }\textbf {\bibinfo
  {volume} {5}},\ \bibinfo {pages} {426} (\bibinfo {year} {2009})}\BibitemShut
  {NoStop}%
\bibitem [{\citenamefont {Ballerini}\ \emph {et~al.}(2008)\citenamefont
  {Ballerini}, \citenamefont {Cabibbo}, \citenamefont {Candelier},
  \citenamefont {Cavagna}, \citenamefont {Cisbani}, \citenamefont {Giardina},
  \citenamefont {Lecomte}, \citenamefont {Orlandi}, \citenamefont {Parisi},
  \citenamefont {Procaccini}, \citenamefont {Viale},\ and\ \citenamefont
  {Zdravkovic}}]{Ballerini2008}%
  \BibitemOpen
  \bibfield  {author} {\bibinfo {author} {\bibfnamefont {M.}~\bibnamefont
  {Ballerini}}, \bibinfo {author} {\bibfnamefont {N.}~\bibnamefont {Cabibbo}},
  \bibinfo {author} {\bibfnamefont {R.}~\bibnamefont {Candelier}}, \bibinfo
  {author} {\bibfnamefont {A.}~\bibnamefont {Cavagna}}, \bibinfo {author}
  {\bibfnamefont {E.}~\bibnamefont {Cisbani}}, \bibinfo {author} {\bibfnamefont
  {I.}~\bibnamefont {Giardina}}, \bibinfo {author} {\bibfnamefont
  {V.}~\bibnamefont {Lecomte}}, \bibinfo {author} {\bibfnamefont
  {A.}~\bibnamefont {Orlandi}}, \bibinfo {author} {\bibfnamefont
  {G.}~\bibnamefont {Parisi}}, \bibinfo {author} {\bibfnamefont
  {A.}~\bibnamefont {Procaccini}}, \bibinfo {author} {\bibfnamefont
  {M.}~\bibnamefont {Viale}}, \ and\ \bibinfo {author} {\bibfnamefont
  {V.}~\bibnamefont {Zdravkovic}},\ }\href@noop {} {\bibfield  {journal}
  {\bibinfo  {journal} {Proceedings of the National Academy of Sciences}\
  }\textbf {\bibinfo {volume} {105}},\ \bibinfo {pages} {1232} (\bibinfo {year}
  {2008})}\BibitemShut {NoStop}%
\bibitem [{\citenamefont {Palacci}\ \emph {et~al.}(2010)\citenamefont
  {Palacci}, \citenamefont {Cottin-Bizonne}, \citenamefont {Ybert},\ and\
  \citenamefont {Bocquet}}]{Palacci2010a}%
  \BibitemOpen
  \bibfield  {author} {\bibinfo {author} {\bibfnamefont {J.}~\bibnamefont
  {Palacci}}, \bibinfo {author} {\bibfnamefont {C.}~\bibnamefont
  {Cottin-Bizonne}}, \bibinfo {author} {\bibfnamefont {C.}~\bibnamefont
  {Ybert}}, \ and\ \bibinfo {author} {\bibfnamefont {L.}~\bibnamefont
  {Bocquet}},\ }\href {\doibase 10.1103/PhysRevLett.105.088304} {\bibfield
  {journal} {\bibinfo  {journal} {Phys. Rev. Lett.}\ }\textbf {\bibinfo
  {volume} {105}},\ \bibinfo {pages} {088304} (\bibinfo {year}
  {2010})}\BibitemShut {NoStop}%
\bibitem [{\citenamefont {Paxton}\ \emph {et~al.}(2004)\citenamefont {Paxton},
  \citenamefont {Kistler}, \citenamefont {Olmeda}, \citenamefont {Sen},
  \citenamefont {St.~Angelo}, \citenamefont {Cao}, \citenamefont {Mallouk},
  \citenamefont {Lammert},\ and\ \citenamefont {Crespi}}]{Paxton2004}%
  \BibitemOpen
  \bibfield  {author} {\bibinfo {author} {\bibfnamefont {W.~F.}\ \bibnamefont
  {Paxton}}, \bibinfo {author} {\bibfnamefont {K.~C.}\ \bibnamefont {Kistler}},
  \bibinfo {author} {\bibfnamefont {C.~C.}\ \bibnamefont {Olmeda}}, \bibinfo
  {author} {\bibfnamefont {A.}~\bibnamefont {Sen}}, \bibinfo {author}
  {\bibfnamefont {S.~K.}\ \bibnamefont {St.~Angelo}}, \bibinfo {author}
  {\bibfnamefont {Y.}~\bibnamefont {Cao}}, \bibinfo {author} {\bibfnamefont
  {T.~E.}\ \bibnamefont {Mallouk}}, \bibinfo {author} {\bibfnamefont {P.~E.}\
  \bibnamefont {Lammert}}, \ and\ \bibinfo {author} {\bibfnamefont {V.~H.}\
  \bibnamefont {Crespi}},\ }\href {\doibase 10.1021/ja047697z} {\bibfield
  {journal} {\bibinfo  {journal} {J. Am. Chem. Soc.}\ }\textbf {\bibinfo
  {volume} {126}},\ \bibinfo {pages} {13424} (\bibinfo {year}
  {2004})}\BibitemShut {NoStop}%
\bibitem [{\citenamefont {Hong}\ \emph {et~al.}(2007)\citenamefont {Hong},
  \citenamefont {Blackman}, \citenamefont {Kopp}, \citenamefont {Sen},\ and\
  \citenamefont {Velegol}}]{Hong2007}%
  \BibitemOpen
  \bibfield  {author} {\bibinfo {author} {\bibfnamefont {Y.}~\bibnamefont
  {Hong}}, \bibinfo {author} {\bibfnamefont {N.~M.~K.}\ \bibnamefont
  {Blackman}}, \bibinfo {author} {\bibfnamefont {N.~D.}\ \bibnamefont {Kopp}},
  \bibinfo {author} {\bibfnamefont {A.}~\bibnamefont {Sen}}, \ and\ \bibinfo
  {author} {\bibfnamefont {D.}~\bibnamefont {Velegol}},\ }\href {\doibase
  10.1103/PhysRevLett.99.178103} {\bibfield  {journal} {\bibinfo  {journal}
  {Phys. Rev. Lett.}\ }\textbf {\bibinfo {volume} {99}},\ \bibinfo {pages}
  {178103} (\bibinfo {year} {2007})}\BibitemShut {NoStop}%
\bibitem [{\citenamefont {Jiang}\ \emph {et~al.}(2010)\citenamefont {Jiang},
  \citenamefont {Yoshinaga},\ and\ \citenamefont {Sano}}]{Jiang2010}%
  \BibitemOpen
  \bibfield  {author} {\bibinfo {author} {\bibfnamefont {H.-R.}\ \bibnamefont
  {Jiang}}, \bibinfo {author} {\bibfnamefont {N.}~\bibnamefont {Yoshinaga}}, \
  and\ \bibinfo {author} {\bibfnamefont {M.}~\bibnamefont {Sano}},\ }\href
  {\doibase 10.1103/PhysRevLett.105.268302} {\bibfield  {journal} {\bibinfo
  {journal} {Phys. Rev. Lett.}\ }\textbf {\bibinfo {volume} {105}},\ \bibinfo
  {pages} {268302} (\bibinfo {year} {2010})}\BibitemShut {NoStop}%
\bibitem [{\citenamefont {Volpe}\ \emph {et~al.}(2011)\citenamefont {Volpe},
  \citenamefont {Buttinoni}, \citenamefont {Vogt}, \citenamefont {Kummerer},\
  and\ \citenamefont {Bechinger}}]{Volpe2011}%
  \BibitemOpen
  \bibfield  {author} {\bibinfo {author} {\bibfnamefont {G.}~\bibnamefont
  {Volpe}}, \bibinfo {author} {\bibfnamefont {I.}~\bibnamefont {Buttinoni}},
  \bibinfo {author} {\bibfnamefont {D.}~\bibnamefont {Vogt}}, \bibinfo {author}
  {\bibfnamefont {H.-J.}\ \bibnamefont {Kummerer}}, \ and\ \bibinfo {author}
  {\bibfnamefont {C.}~\bibnamefont {Bechinger}},\ }\href
  {http://dx.doi.org/10.1039/C1SM05960B} {\bibfield  {journal} {\bibinfo
  {journal} {Soft Matter}\ }\textbf {\bibinfo {volume} {7}},\ \bibinfo {pages}
  {8810} (\bibinfo {year} {2011})}\BibitemShut {NoStop}%
\bibitem [{\citenamefont {Thutupalli}\ \emph {et~al.}(2011)\citenamefont
  {Thutupalli}, \citenamefont {Seemann},\ and\ \citenamefont
  {Herminghaus}}]{Thutupalli2011}%
  \BibitemOpen
  \bibfield  {author} {\bibinfo {author} {\bibfnamefont {S.}~\bibnamefont
  {Thutupalli}}, \bibinfo {author} {\bibfnamefont {R.}~\bibnamefont {Seemann}},
  \ and\ \bibinfo {author} {\bibfnamefont {S.}~\bibnamefont {Herminghaus}},\
  }\href {\doibase 10.1088/1367-2630/13/7/073021} {\bibfield  {journal}
  {\bibinfo  {journal} {New Journal of Physics}\ }\textbf {\bibinfo {volume}
  {13}},\ \bibinfo {pages} {073021} (\bibinfo {year} {2011})}\BibitemShut
  {NoStop}%
\bibitem [{\citenamefont {Bricard}\ \emph {et~al.}(2013)\citenamefont
  {Bricard}, \citenamefont {Caussin}, \citenamefont {Desreumaux}, \citenamefont
  {Dauchot},\ and\ \citenamefont {Bartolo}}]{Bricard2013}%
  \BibitemOpen
  \bibfield  {author} {\bibinfo {author} {\bibfnamefont {A.}~\bibnamefont
  {Bricard}}, \bibinfo {author} {\bibfnamefont {J.-B.}\ \bibnamefont
  {Caussin}}, \bibinfo {author} {\bibfnamefont {N.}~\bibnamefont {Desreumaux}},
  \bibinfo {author} {\bibfnamefont {O.}~\bibnamefont {Dauchot}}, \ and\
  \bibinfo {author} {\bibfnamefont {D.}~\bibnamefont {Bartolo}},\ }\href
  {\doibase 10.1038/nature12673} {\bibfield  {journal} {\bibinfo  {journal}
  {Nature}\ }\textbf {\bibinfo {volume} {503}},\ \bibinfo {pages} {95}
  (\bibinfo {year} {2013})}\BibitemShut {NoStop}%
\bibitem [{\citenamefont {Narayan}\ \emph {et~al.}(2007)\citenamefont
  {Narayan}, \citenamefont {Ramaswamy},\ and\ \citenamefont
  {Menon}}]{Narayan2007}%
  \BibitemOpen
  \bibfield  {author} {\bibinfo {author} {\bibfnamefont {V.}~\bibnamefont
  {Narayan}}, \bibinfo {author} {\bibfnamefont {S.}~\bibnamefont {Ramaswamy}},
  \ and\ \bibinfo {author} {\bibfnamefont {N.}~\bibnamefont {Menon}},\ }\href
  {\doibase 10.1126/science.1140414} {\bibfield  {journal} {\bibinfo  {journal}
  {Science}\ }\textbf {\bibinfo {volume} {317}},\ \bibinfo {pages} {105}
  (\bibinfo {year} {2007})}\BibitemShut {NoStop}%
\bibitem [{\citenamefont {Kudrolli}\ \emph {et~al.}(2008)\citenamefont
  {Kudrolli}, \citenamefont {Lumay}, \citenamefont {Volfson},\ and\
  \citenamefont {Tsimring}}]{Kudrolli2008}%
  \BibitemOpen
  \bibfield  {author} {\bibinfo {author} {\bibfnamefont {A.}~\bibnamefont
  {Kudrolli}}, \bibinfo {author} {\bibfnamefont {G.}~\bibnamefont {Lumay}},
  \bibinfo {author} {\bibfnamefont {D.}~\bibnamefont {Volfson}}, \ and\
  \bibinfo {author} {\bibfnamefont {L.~S.}\ \bibnamefont {Tsimring}},\ }\href
  {\doibase 10.1103/PhysRevLett.100.058001} {\bibfield  {journal} {\bibinfo
  {journal} {Phys. Rev. Lett.}\ }\textbf {\bibinfo {volume} {100}},\ \bibinfo
  {pages} {058001} (\bibinfo {year} {2008})}\BibitemShut {NoStop}%
\bibitem [{\citenamefont {Deseigne}\ \emph {et~al.}(2010)\citenamefont
  {Deseigne}, \citenamefont {Dauchot},\ and\ \citenamefont
  {Chat\'e}}]{Deseigne2010}%
  \BibitemOpen
  \bibfield  {author} {\bibinfo {author} {\bibfnamefont {J.}~\bibnamefont
  {Deseigne}}, \bibinfo {author} {\bibfnamefont {O.}~\bibnamefont {Dauchot}}, \
  and\ \bibinfo {author} {\bibfnamefont {H.}~\bibnamefont {Chat\'e}},\ }\href
  {\doibase 10.1103/PhysRevLett.105.098001} {\bibfield  {journal} {\bibinfo
  {journal} {Phys. Rev. Lett.}\ }\textbf {\bibinfo {volume} {105}},\ \bibinfo
  {pages} {098001} (\bibinfo {year} {2010})}\BibitemShut {NoStop}%
\bibitem [{\citenamefont {Angelani}\ \emph {et~al.}(2009)\citenamefont
  {Angelani}, \citenamefont {Di~Leonardo},\ and\ \citenamefont
  {Ruocco}}]{Angelani2009}%
  \BibitemOpen
  \bibfield  {author} {\bibinfo {author} {\bibfnamefont {L.}~\bibnamefont
  {Angelani}}, \bibinfo {author} {\bibfnamefont {R.}~\bibnamefont
  {Di~Leonardo}}, \ and\ \bibinfo {author} {\bibfnamefont {G.}~\bibnamefont
  {Ruocco}},\ }\href {\doibase 10.1103/PhysRevLett.102.048104} {\bibfield
  {journal} {\bibinfo  {journal} {Phys. Rev. Lett.}\ }\textbf {\bibinfo
  {volume} {102}},\ \bibinfo {pages} {048104} (\bibinfo {year}
  {2009})}\BibitemShut {NoStop}%
\bibitem [{\citenamefont {Di~Leonardo}\ \emph {et~al.}(2010)\citenamefont
  {Di~Leonardo}, \citenamefont {Angelani}, \citenamefont {Dell'Arciprete},
  \citenamefont {Ruocco}, \citenamefont {Iebba}, \citenamefont {Schippa},
  \citenamefont {Conte}, \citenamefont {Mecarini}, \citenamefont {De~Angelis},\
  and\ \citenamefont {Di~Fabrizio}}]{DiLeonardo2010}%
  \BibitemOpen
  \bibfield  {author} {\bibinfo {author} {\bibfnamefont {R.}~\bibnamefont
  {Di~Leonardo}}, \bibinfo {author} {\bibfnamefont {L.}~\bibnamefont
  {Angelani}}, \bibinfo {author} {\bibfnamefont {D.}~\bibnamefont
  {Dell'Arciprete}}, \bibinfo {author} {\bibfnamefont {G.}~\bibnamefont
  {Ruocco}}, \bibinfo {author} {\bibfnamefont {V.}~\bibnamefont {Iebba}},
  \bibinfo {author} {\bibfnamefont {S.}~\bibnamefont {Schippa}}, \bibinfo
  {author} {\bibfnamefont {M.~P.}\ \bibnamefont {Conte}}, \bibinfo {author}
  {\bibfnamefont {F.}~\bibnamefont {Mecarini}}, \bibinfo {author}
  {\bibfnamefont {F.}~\bibnamefont {De~Angelis}}, \ and\ \bibinfo {author}
  {\bibfnamefont {E.}~\bibnamefont {Di~Fabrizio}},\ }\href {\doibase
  10.1073/pnas.0910426107} {\bibfield  {journal} {\bibinfo  {journal}
  {Proceedings of the National Academy of Sciences}\ }\textbf {\bibinfo
  {volume} {107}},\ \bibinfo {pages} {9541} (\bibinfo {year}
  {2010})}\BibitemShut {NoStop}%
\bibitem [{\citenamefont {Sokolov}\ \emph {et~al.}(2010)\citenamefont
  {Sokolov}, \citenamefont {Apodaca}, \citenamefont {Grzybowski},\ and\
  \citenamefont {Aranson}}]{Sokolov2010}%
  \BibitemOpen
  \bibfield  {author} {\bibinfo {author} {\bibfnamefont {A.}~\bibnamefont
  {Sokolov}}, \bibinfo {author} {\bibfnamefont {M.~M.}\ \bibnamefont
  {Apodaca}}, \bibinfo {author} {\bibfnamefont {B.~A.}\ \bibnamefont
  {Grzybowski}}, \ and\ \bibinfo {author} {\bibfnamefont {I.~S.}\ \bibnamefont
  {Aranson}},\ }\href {\doibase 10.1073/pnas.0913015107} {\bibfield  {journal}
  {\bibinfo  {journal} {Proceedings of the National Academy of Sciences}\
  }\textbf {\bibinfo {volume} {107}},\ \bibinfo {pages} {969} (\bibinfo {year}
  {2010})}\BibitemShut {NoStop}%
\bibitem [{\citenamefont {Galajda}\ \emph {et~al.}(2007)\citenamefont
  {Galajda}, \citenamefont {Keymer}, \citenamefont {Chaikin},\ and\
  \citenamefont {Austin}}]{Galajda2007}%
  \BibitemOpen
  \bibfield  {author} {\bibinfo {author} {\bibfnamefont {P.}~\bibnamefont
  {Galajda}}, \bibinfo {author} {\bibfnamefont {J.}~\bibnamefont {Keymer}},
  \bibinfo {author} {\bibfnamefont {P.}~\bibnamefont {Chaikin}}, \ and\
  \bibinfo {author} {\bibfnamefont {R.}~\bibnamefont {Austin}},\ }\href
  {\doibase 10.1128/JB.01033-07} {\bibfield  {journal} {\bibinfo  {journal}
  {Journal of Bacteriology}\ }\textbf {\bibinfo {volume} {189}},\ \bibinfo
  {pages} {8704} (\bibinfo {year} {2007})}\BibitemShut {NoStop}%
\bibitem [{\citenamefont {Wan}\ \emph {et~al.}(2008)\citenamefont {Wan},
  \citenamefont {Olson~Reichhardt}, \citenamefont {Nussinov},\ and\
  \citenamefont {Reichhardt}}]{Wan2008}%
  \BibitemOpen
  \bibfield  {author} {\bibinfo {author} {\bibfnamefont {M.~B.}\ \bibnamefont
  {Wan}}, \bibinfo {author} {\bibfnamefont {C.~J.}\ \bibnamefont
  {Olson~Reichhardt}}, \bibinfo {author} {\bibfnamefont {Z.}~\bibnamefont
  {Nussinov}}, \ and\ \bibinfo {author} {\bibfnamefont {C.}~\bibnamefont
  {Reichhardt}},\ }\href {\doibase 10.1103/PhysRevLett.101.018102} {\bibfield
  {journal} {\bibinfo  {journal} {Phys. Rev. Lett.}\ }\textbf {\bibinfo
  {volume} {101}},\ \bibinfo {pages} {018102} (\bibinfo {year}
  {2008})}\BibitemShut {NoStop}%
\bibitem [{\citenamefont {Tailleur}\ and\ \citenamefont
  {Cates}(2009)}]{Tailleur2009}%
  \BibitemOpen
  \bibfield  {author} {\bibinfo {author} {\bibfnamefont {J.}~\bibnamefont
  {Tailleur}}\ and\ \bibinfo {author} {\bibfnamefont {M.~E.}\ \bibnamefont
  {Cates}},\ }\href {http://stacks.iop.org/0295-5075/86/i=6/a=60002} {\bibfield
   {journal} {\bibinfo  {journal} {EPL (Europhysics Letters)}\ }\textbf
  {\bibinfo {volume} {86}},\ \bibinfo {pages} {60002} (\bibinfo {year}
  {2009})}\BibitemShut {NoStop}%
\bibitem [{\citenamefont {Angelani}\ \emph {et~al.}(2011)\citenamefont
  {Angelani}, \citenamefont {Costanzo},\ and\ \citenamefont
  {Leonardo}}]{Angelani2011}%
  \BibitemOpen
  \bibfield  {author} {\bibinfo {author} {\bibfnamefont {L.}~\bibnamefont
  {Angelani}}, \bibinfo {author} {\bibfnamefont {A.}~\bibnamefont {Costanzo}},
  \ and\ \bibinfo {author} {\bibfnamefont {R.~D.}\ \bibnamefont {Leonardo}},\
  }\href {\doibase 10.1209/0295-5075/96/68002} {\bibfield  {journal} {\bibinfo
  {journal} {EPL (Europhysics Letters)}\ }\textbf {\bibinfo {volume} {96}},\
  \bibinfo {pages} {68002} (\bibinfo {year} {2011})}\BibitemShut {NoStop}%
\bibitem [{\citenamefont {Ghosh}\ \emph {et~al.}(2013)\citenamefont {Ghosh},
  \citenamefont {Misko}, \citenamefont {Marchesoni},\ and\ \citenamefont
  {Nori}}]{Ghosh2013}%
  \BibitemOpen
  \bibfield  {author} {\bibinfo {author} {\bibfnamefont {P.~K.}\ \bibnamefont
  {Ghosh}}, \bibinfo {author} {\bibfnamefont {V.~R.}\ \bibnamefont {Misko}},
  \bibinfo {author} {\bibfnamefont {F.}~\bibnamefont {Marchesoni}}, \ and\
  \bibinfo {author} {\bibfnamefont {F.}~\bibnamefont {Nori}},\ }\href {\doibase
  10.1103/PhysRevLett.110.268301} {\bibfield  {journal} {\bibinfo  {journal}
  {Phys. Rev. Lett.}\ }\textbf {\bibinfo {volume} {110}},\ \bibinfo {pages}
  {268301} (\bibinfo {year} {2013})}\BibitemShut {NoStop}%
\bibitem [{\citenamefont {Ai}\ \emph {et~al.}(2013)\citenamefont {Ai},
  \citenamefont {Chen}, \citenamefont {He}, \citenamefont {Li},\ and\
  \citenamefont {Zhong}}]{Ai2013}%
  \BibitemOpen
  \bibfield  {author} {\bibinfo {author} {\bibfnamefont {B.-q.}\ \bibnamefont
  {Ai}}, \bibinfo {author} {\bibfnamefont {Q.-y.}\ \bibnamefont {Chen}},
  \bibinfo {author} {\bibfnamefont {Y.-f.}\ \bibnamefont {He}}, \bibinfo
  {author} {\bibfnamefont {F.-g.}\ \bibnamefont {Li}}, \ and\ \bibinfo {author}
  {\bibfnamefont {W.-r.}\ \bibnamefont {Zhong}},\ }\href {\doibase
  10.1103/PhysRevE.88.062129} {\bibfield  {journal} {\bibinfo  {journal} {Phys.
  Rev. E}\ }\textbf {\bibinfo {volume} {88}},\ \bibinfo {pages} {062129}
  (\bibinfo {year} {2013})}\BibitemShut {NoStop}%
\bibitem [{\citenamefont {Nash}\ \emph {et~al.}(2010)\citenamefont {Nash},
  \citenamefont {Adhikari}, \citenamefont {Tailleur},\ and\ \citenamefont
  {Cates}}]{Nash2010}%
  \BibitemOpen
  \bibfield  {author} {\bibinfo {author} {\bibfnamefont {R.~W.}\ \bibnamefont
  {Nash}}, \bibinfo {author} {\bibfnamefont {R.}~\bibnamefont {Adhikari}},
  \bibinfo {author} {\bibfnamefont {J.}~\bibnamefont {Tailleur}}, \ and\
  \bibinfo {author} {\bibfnamefont {M.~E.}\ \bibnamefont {Cates}},\ }\href
  {http://link.aps.org/doi/10.1103/PhysRevLett.104.258101} {\bibfield
  {journal} {\bibinfo  {journal} {Phys. Rev. Lett.}\ }\textbf {\bibinfo
  {volume} {104}},\ \bibinfo {pages} {258101} (\bibinfo {year}
  {2010})}\BibitemShut {NoStop}%
\bibitem [{\citenamefont {Elgeti}\ and\ \citenamefont
  {Gompper}(2013)}]{Elgeti2013}%
  \BibitemOpen
  \bibfield  {author} {\bibinfo {author} {\bibfnamefont {J.}~\bibnamefont
  {Elgeti}}\ and\ \bibinfo {author} {\bibfnamefont {G.}~\bibnamefont
  {Gompper}},\ }\href {http://stacks.iop.org/0295-5075/101/i=4/a=48003}
  {\bibfield  {journal} {\bibinfo  {journal} {EPL (Europhysics Letters)}\
  }\textbf {\bibinfo {volume} {101}},\ \bibinfo {pages} {48003} (\bibinfo
  {year} {2013})}\BibitemShut {NoStop}%
\bibitem [{\citenamefont {Lee}(2013)}]{Lee2013}%
  \BibitemOpen
  \bibfield  {author} {\bibinfo {author} {\bibfnamefont {C.~F.}\ \bibnamefont
  {Lee}},\ }\href {http://stacks.iop.org/1367-2630/15/i=5/a=055007} {\bibfield
  {journal} {\bibinfo  {journal} {New Journal of Physics}\ }\textbf {\bibinfo
  {volume} {15}},\ \bibinfo {pages} {055007} (\bibinfo {year}
  {2013})}\BibitemShut {NoStop}%
\bibitem [{\citenamefont {Mallory}\ \emph {et~al.}(2014)\citenamefont
  {Mallory}, \citenamefont {\v{S}ari\'c}, \citenamefont {Valeriani},\ and\
  \citenamefont {Cacciuto}}]{Mallory2014}%
  \BibitemOpen
  \bibfield  {author} {\bibinfo {author} {\bibfnamefont {S.~A.}\ \bibnamefont
  {Mallory}}, \bibinfo {author} {\bibfnamefont {A.}~\bibnamefont
  {\v{S}ari\'c}}, \bibinfo {author} {\bibfnamefont {C.}~\bibnamefont
  {Valeriani}}, \ and\ \bibinfo {author} {\bibfnamefont {A.}~\bibnamefont
  {Cacciuto}},\ }\href {http://link.aps.org/doi/10.1103/PhysRevE.89.052303}
  {\bibfield  {journal} {\bibinfo  {journal} {Phys. Rev. E}\ }\textbf {\bibinfo
  {volume} {89}},\ \bibinfo {pages} {052303} (\bibinfo {year}
  {2014})}\BibitemShut {NoStop}%
\bibitem [{\citenamefont {Kaiser}\ \emph {et~al.}(2012)\citenamefont {Kaiser},
  \citenamefont {Wensink},\ and\ \citenamefont {L\"owen}}]{Kaiser2012}%
  \BibitemOpen
  \bibfield  {author} {\bibinfo {author} {\bibfnamefont {A.}~\bibnamefont
  {Kaiser}}, \bibinfo {author} {\bibfnamefont {H.~H.}\ \bibnamefont {Wensink}},
  \ and\ \bibinfo {author} {\bibfnamefont {H.}~\bibnamefont {L\"owen}},\ }\href
  {\doibase 10.1103/PhysRevLett.108.268307} {\bibfield  {journal} {\bibinfo
  {journal} {Phys. Rev. Lett.}\ }\textbf {\bibinfo {volume} {108}},\ \bibinfo
  {pages} {268307} (\bibinfo {year} {2012})}\BibitemShut {NoStop}%
\bibitem [{\citenamefont {Kaiser}\ \emph {et~al.}(2013)\citenamefont {Kaiser},
  \citenamefont {Popowa}, \citenamefont {Wensink},\ and\ \citenamefont
  {L\"owen}}]{Kaiser2013}%
  \BibitemOpen
  \bibfield  {author} {\bibinfo {author} {\bibfnamefont {A.}~\bibnamefont
  {Kaiser}}, \bibinfo {author} {\bibfnamefont {K.}~\bibnamefont {Popowa}},
  \bibinfo {author} {\bibfnamefont {H.~H.}\ \bibnamefont {Wensink}}, \ and\
  \bibinfo {author} {\bibfnamefont {H.}~\bibnamefont {L\"owen}},\ }\href
  {\doibase 10.1103/PhysRevE.88.022311} {\bibfield  {journal} {\bibinfo
  {journal} {Phys. Rev. E}\ }\textbf {\bibinfo {volume} {88}},\ \bibinfo
  {pages} {022311} (\bibinfo {year} {2013})}\BibitemShut {NoStop}%
\bibitem [{\citenamefont {Wioland}\ \emph {et~al.}(2013)\citenamefont
  {Wioland}, \citenamefont {Woodhouse}, \citenamefont {Dunkel}, \citenamefont
  {Kessler},\ and\ \citenamefont {Goldstein}}]{Wioland2013}%
  \BibitemOpen
  \bibfield  {author} {\bibinfo {author} {\bibfnamefont {H.}~\bibnamefont
  {Wioland}}, \bibinfo {author} {\bibfnamefont {F.~G.}\ \bibnamefont
  {Woodhouse}}, \bibinfo {author} {\bibfnamefont {J.}~\bibnamefont {Dunkel}},
  \bibinfo {author} {\bibfnamefont {J.~O.}\ \bibnamefont {Kessler}}, \ and\
  \bibinfo {author} {\bibfnamefont {R.~E.}\ \bibnamefont {Goldstein}},\ }\href
  {http://link.aps.org/doi/10.1103/PhysRevLett.110.268102} {\bibfield
  {journal} {\bibinfo  {journal} {Phys. Rev. Lett.}\ }\textbf {\bibinfo
  {volume} {110}},\ \bibinfo {pages} {268102} (\bibinfo {year}
  {2013})}\BibitemShut {NoStop}%
\bibitem [{\citenamefont {Kantsler}\ \emph {et~al.}(2013)\citenamefont
  {Kantsler}, \citenamefont {Dunkel}, \citenamefont {Polin},\ and\
  \citenamefont {Goldstein}}]{Kantsler2013}%
  \BibitemOpen
  \bibfield  {author} {\bibinfo {author} {\bibfnamefont {V.}~\bibnamefont
  {Kantsler}}, \bibinfo {author} {\bibfnamefont {J.}~\bibnamefont {Dunkel}},
  \bibinfo {author} {\bibfnamefont {M.}~\bibnamefont {Polin}}, \ and\ \bibinfo
  {author} {\bibfnamefont {R.~E.}\ \bibnamefont {Goldstein}},\ }\href {\doibase
  10.1073/pnas.1210548110} {\bibfield  {journal} {\bibinfo  {journal}
  {Proceedings of the National Academy of Sciences}\ }\textbf {\bibinfo
  {volume} {110}},\ \bibinfo {pages} {1187} (\bibinfo {year}
  {2013})}\BibitemShut {NoStop}%
\bibitem [{\citenamefont {Guidobaldi}\ \emph {et~al.}(2014)\citenamefont
  {Guidobaldi}, \citenamefont {Jeyaram}, \citenamefont {Berdakin},
  \citenamefont {Moshchalkov}, \citenamefont {Condat}, \citenamefont {Marconi},
  \citenamefont {Giojalas},\ and\ \citenamefont {Silhanek}}]{Guidobaldi2014}%
  \BibitemOpen
  \bibfield  {author} {\bibinfo {author} {\bibfnamefont {A.}~\bibnamefont
  {Guidobaldi}}, \bibinfo {author} {\bibfnamefont {Y.}~\bibnamefont {Jeyaram}},
  \bibinfo {author} {\bibfnamefont {I.}~\bibnamefont {Berdakin}}, \bibinfo
  {author} {\bibfnamefont {V.~V.}\ \bibnamefont {Moshchalkov}}, \bibinfo
  {author} {\bibfnamefont {C.~A.}\ \bibnamefont {Condat}}, \bibinfo {author}
  {\bibfnamefont {V.~I.}\ \bibnamefont {Marconi}}, \bibinfo {author}
  {\bibfnamefont {L.}~\bibnamefont {Giojalas}}, \ and\ \bibinfo {author}
  {\bibfnamefont {A.~V.}\ \bibnamefont {Silhanek}},\ }\href {\doibase
  10.1103/PhysRevE.89.032720} {\bibfield  {journal} {\bibinfo  {journal} {Phys.
  Rev. E}\ }\textbf {\bibinfo {volume} {89}},\ \bibinfo {pages} {032720}
  (\bibinfo {year} {2014})}\BibitemShut {NoStop}%
\bibitem [{\citenamefont {Yang}\ \emph {et~al.}(2014)\citenamefont {Yang},
  \citenamefont {Manning},\ and\ \citenamefont {Marchetti}}]{Yang2014}%
  \BibitemOpen
  \bibfield  {author} {\bibinfo {author} {\bibfnamefont {X.}~\bibnamefont
  {Yang}}, \bibinfo {author} {\bibfnamefont {M.~L.}\ \bibnamefont {Manning}}, \
  and\ \bibinfo {author} {\bibfnamefont {M.~C.}\ \bibnamefont {Marchetti}},\
  }\href {http://arxiv.org/abs/1308.3741} {\enquote {\bibinfo {title}
  {Aggregation and segregation of confined active particles},}\ }\bibinfo
  {howpublished} {arXiv:1403.0697} (\bibinfo {year} {2014})\BibitemShut
  {NoStop}%
\bibitem [{\citenamefont {Camley}\ and\ \citenamefont {Rappel}()}]{Camley2014}%
  \BibitemOpen
  \bibfield  {author} {\bibinfo {author} {\bibfnamefont {B.~A.}\ \bibnamefont
  {Camley}}\ and\ \bibinfo {author} {\bibfnamefont {W.-J.}\ \bibnamefont
  {Rappel}},\ }\href {http://arxiv.org/abs/1405.7088} {\enquote {\bibinfo
  {title} {Velocity alignment leads to high persistence in confined cells},}\
  }\bibinfo {howpublished} {arXiv:1405.7088}\BibitemShut {NoStop}%
\bibitem [{\citenamefont {Ray}\ \emph {et~al.}(2014)\citenamefont {Ray},
  \citenamefont {Reichhardt},\ and\ \citenamefont {Reichhardt}}]{Ray2014}%
  \BibitemOpen
  \bibfield  {author} {\bibinfo {author} {\bibfnamefont {D.}~\bibnamefont
  {Ray}}, \bibinfo {author} {\bibfnamefont {C.}~\bibnamefont {Reichhardt}}, \
  and\ \bibinfo {author} {\bibfnamefont {C.~J.~O.}\ \bibnamefont
  {Reichhardt}},\ }\href {http://link.aps.org/doi/10.1103/PhysRevE.90.013019}
  {\bibfield  {journal} {\bibinfo  {journal} {Phys. Rev. E}\ }\textbf {\bibinfo
  {volume} {90}},\ \bibinfo {pages} {013019} (\bibinfo {year}
  {2014})}\BibitemShut {NoStop}%
\bibitem [{\citenamefont {Lushi}\ \emph {et~al.}(2014)\citenamefont {Lushi},
  \citenamefont {Wioland},\ and\ \citenamefont {Goldstein}}]{Lushi2014}%
  \BibitemOpen
  \bibfield  {author} {\bibinfo {author} {\bibfnamefont {E.}~\bibnamefont
  {Lushi}}, \bibinfo {author} {\bibfnamefont {H.}~\bibnamefont {Wioland}}, \
  and\ \bibinfo {author} {\bibfnamefont {R.~E.}\ \bibnamefont {Goldstein}},\
  }\href {\doibase 10.1073/pnas.1405698111} {\bibfield  {journal} {\bibinfo
  {journal} {Proceedings of the National Academy of Sciences}\ }\textbf
  {\bibinfo {volume} {111}},\ \bibinfo {pages} {9733} (\bibinfo {year}
  {2014})}\BibitemShut {NoStop}%
\bibitem [{\citenamefont {Kaiser}\ \emph {et~al.}(2014)\citenamefont {Kaiser},
  \citenamefont {Peshkov}, \citenamefont {Sokolov}, \citenamefont {ten Hagen},
  \citenamefont {Löwen},\ and\ \citenamefont {Aranson}}]{Kaiser2014}%
  \BibitemOpen
  \bibfield  {author} {\bibinfo {author} {\bibfnamefont {A.}~\bibnamefont
  {Kaiser}}, \bibinfo {author} {\bibfnamefont {A.}~\bibnamefont {Peshkov}},
  \bibinfo {author} {\bibfnamefont {A.}~\bibnamefont {Sokolov}}, \bibinfo
  {author} {\bibfnamefont {B.}~\bibnamefont {ten Hagen}}, \bibinfo {author}
  {\bibfnamefont {H.}~\bibnamefont {Löwen}}, \ and\ \bibinfo {author}
  {\bibfnamefont {I.~S.}\ \bibnamefont {Aranson}},\ }\href
  {http://link.aps.org/doi/10.1103/PhysRevLett.112.158101} {\bibfield
  {journal} {\bibinfo  {journal} {Phys. Rev. Lett.}\ }\textbf {\bibinfo
  {volume} {112}},\ \bibinfo {pages} {158101} (\bibinfo {year}
  {2014})}\BibitemShut {NoStop}%
\bibitem [{\citenamefont {Fily}\ and\ \citenamefont
  {Marchetti}(2012)}]{Fily2012}%
  \BibitemOpen
  \bibfield  {author} {\bibinfo {author} {\bibfnamefont {Y.}~\bibnamefont
  {Fily}}\ and\ \bibinfo {author} {\bibfnamefont {M.~C.}\ \bibnamefont
  {Marchetti}},\ }\href {\doibase 10.1103/PhysRevLett.108.235702} {\bibfield
  {journal} {\bibinfo  {journal} {Physical Review Letters}\ }\textbf {\bibinfo
  {volume} {108}},\ \bibinfo {pages} {235702} (\bibinfo {year}
  {2012})}\BibitemShut {NoStop}%
\bibitem [{\citenamefont {Redner}\ \emph {et~al.}(2013)\citenamefont {Redner},
  \citenamefont {Hagan},\ and\ \citenamefont {Baskaran}}]{Redner2013}%
  \BibitemOpen
  \bibfield  {author} {\bibinfo {author} {\bibfnamefont {G.~S.}\ \bibnamefont
  {Redner}}, \bibinfo {author} {\bibfnamefont {M.~F.}\ \bibnamefont {Hagan}}, \
  and\ \bibinfo {author} {\bibfnamefont {A.}~\bibnamefont {Baskaran}},\ }\href
  {http://link.aps.org/doi/10.1103/PhysRevLett.110.055701} {\bibfield
  {journal} {\bibinfo  {journal} {Phys. Rev. Lett.}\ }\textbf {\bibinfo
  {volume} {110}},\ \bibinfo {pages} {055701} (\bibinfo {year}
  {2013})}\BibitemShut {NoStop}%
\bibitem [{\citenamefont {Bialk\'e}\ \emph {et~al.}(2013)\citenamefont
  {Bialk\'e}, \citenamefont {L\"owen},\ and\ \citenamefont
  {Speck}}]{Bialke2013}%
  \BibitemOpen
  \bibfield  {author} {\bibinfo {author} {\bibfnamefont {J.}~\bibnamefont
  {Bialk\'e}}, \bibinfo {author} {\bibfnamefont {H.}~\bibnamefont {L\"owen}}, \
  and\ \bibinfo {author} {\bibfnamefont {T.}~\bibnamefont {Speck}},\ }\href
  {http://stacks.iop.org/0295-5075/103/i=3/a=30008} {\bibfield  {journal}
  {\bibinfo  {journal} {EPL (Europhysics Letters)}\ }\textbf {\bibinfo {volume}
  {103}},\ \bibinfo {pages} {30008} (\bibinfo {year} {2013})}\BibitemShut
  {NoStop}%
\bibitem [{\citenamefont {Cates}\ and\ \citenamefont
  {Tailleur}(2013)}]{Cates2013}%
  \BibitemOpen
  \bibfield  {author} {\bibinfo {author} {\bibfnamefont {M.~E.}\ \bibnamefont
  {Cates}}\ and\ \bibinfo {author} {\bibfnamefont {J.}~\bibnamefont
  {Tailleur}},\ }\href@noop {} {\bibfield  {journal} {\bibinfo  {journal} {EPL
  (Europhysics Letters)}\ }\textbf {\bibinfo {volume} {101}},\ \bibinfo {pages}
  {20010} (\bibinfo {year} {2013})}\BibitemShut {NoStop}%
\bibitem [{\citenamefont {Stenhammar}\ \emph {et~al.}(2013)\citenamefont
  {Stenhammar}, \citenamefont {Tiribocchi}, \citenamefont {Allen},
  \citenamefont {Marenduzzo},\ and\ \citenamefont {Cates}}]{Stenhammar2013}%
  \BibitemOpen
  \bibfield  {author} {\bibinfo {author} {\bibfnamefont {J.}~\bibnamefont
  {Stenhammar}}, \bibinfo {author} {\bibfnamefont {A.}~\bibnamefont
  {Tiribocchi}}, \bibinfo {author} {\bibfnamefont {R.~J.}\ \bibnamefont
  {Allen}}, \bibinfo {author} {\bibfnamefont {D.}~\bibnamefont {Marenduzzo}}, \
  and\ \bibinfo {author} {\bibfnamefont {M.~E.}\ \bibnamefont {Cates}},\ }\href
  {http://link.aps.org/doi/10.1103/PhysRevLett.111.145702} {\bibfield
  {journal} {\bibinfo  {journal} {Phys. Rev. Lett.}\ }\textbf {\bibinfo
  {volume} {111}},\ \bibinfo {pages} {145702} (\bibinfo {year}
  {2013})}\BibitemShut {NoStop}%
\bibitem [{\citenamefont {Fily}\ \emph
  {et~al.}(2014{\natexlab{a}})\citenamefont {Fily}, \citenamefont {Henkes},\
  and\ \citenamefont {Marchetti}}]{Fily2014}%
  \BibitemOpen
  \bibfield  {author} {\bibinfo {author} {\bibfnamefont {Y.}~\bibnamefont
  {Fily}}, \bibinfo {author} {\bibfnamefont {S.}~\bibnamefont {Henkes}}, \ and\
  \bibinfo {author} {\bibfnamefont {M.~C.}\ \bibnamefont {Marchetti}},\ }\href
  {\doibase 10.1039/C3SM52469H} {\bibfield  {journal} {\bibinfo  {journal}
  {Soft Matter}\ }\textbf {\bibinfo {volume} {10}},\ \bibinfo {pages} {2132}
  (\bibinfo {year} {2014}{\natexlab{a}})}\BibitemShut {NoStop}%
\bibitem [{\citenamefont {Stenhammar}\ \emph {et~al.}(2014)\citenamefont
  {Stenhammar}, \citenamefont {Marenduzzo}, \citenamefont {Allen},\ and\
  \citenamefont {Cates}}]{Stenhammar2014}%
  \BibitemOpen
  \bibfield  {author} {\bibinfo {author} {\bibfnamefont {J.}~\bibnamefont
  {Stenhammar}}, \bibinfo {author} {\bibfnamefont {D.}~\bibnamefont
  {Marenduzzo}}, \bibinfo {author} {\bibfnamefont {R.~J.}\ \bibnamefont
  {Allen}}, \ and\ \bibinfo {author} {\bibfnamefont {M.~E.}\ \bibnamefont
  {Cates}},\ }\href {http://dx.doi.org/10.1039/C3SM52813H} {\bibfield
  {journal} {\bibinfo  {journal} {Soft Matter}\ }\textbf {\bibinfo {volume}
  {10}},\ \bibinfo {pages} {1489} (\bibinfo {year} {2014})}\BibitemShut
  {NoStop}%
\bibitem [{\citenamefont {Wittkowski}\ \emph {et~al.}(2014)\citenamefont
  {Wittkowski}, \citenamefont {Tiribocchi}, \citenamefont {Stenhammar},
  \citenamefont {Allen}, \citenamefont {Marenduzzo},\ and\ \citenamefont
  {Cates}}]{Wittkowski2014}%
  \BibitemOpen
  \bibfield  {author} {\bibinfo {author} {\bibfnamefont {R.}~\bibnamefont
  {Wittkowski}}, \bibinfo {author} {\bibfnamefont {A.}~\bibnamefont
  {Tiribocchi}}, \bibinfo {author} {\bibfnamefont {J.}~\bibnamefont
  {Stenhammar}}, \bibinfo {author} {\bibfnamefont {R.~J.}\ \bibnamefont
  {Allen}}, \bibinfo {author} {\bibfnamefont {D.}~\bibnamefont {Marenduzzo}}, \
  and\ \bibinfo {author} {\bibfnamefont {M.~E.}\ \bibnamefont {Cates}},\ }\href
  {http://arxiv.org/abs/1311.1256} {\enquote {\bibinfo {title} {Scalar $\phi^4$
  field theories for active-particle phase separation},}\ }\bibinfo
  {howpublished} {arXiv:1311.1256} (\bibinfo {year} {2014})\BibitemShut
  {NoStop}%
\bibitem [{\citenamefont {Fily}\ \emph
  {et~al.}(2014{\natexlab{b}})\citenamefont {Fily}, \citenamefont {Baskaran},\
  and\ \citenamefont {Hagan}}]{Fily2014a}%
  \BibitemOpen
  \bibfield  {author} {\bibinfo {author} {\bibfnamefont {Y.}~\bibnamefont
  {Fily}}, \bibinfo {author} {\bibfnamefont {A.}~\bibnamefont {Baskaran}}, \
  and\ \bibinfo {author} {\bibfnamefont {M.~F.}\ \bibnamefont {Hagan}},\ }\href
  {\doibase 10.1039/C4SM00975D} {\bibfield  {journal} {\bibinfo  {journal}
  {Soft Matter}\ }\textbf {\bibinfo {volume} {10}},\ \bibinfo {pages} {5609}
  (\bibinfo {year} {2014}{\natexlab{b}})}\BibitemShut {NoStop}%
\bibitem [{Note1()}]{Note1}%
  \BibitemOpen
  \bibinfo {note} {With this choice of potential, the dynamics does not depend
  on the value of the mobility $\mu $ which is only kept for dimensional
  consistency.}\BibitemShut {Stop}%
\bibitem [{Note2()}]{Note2}%
  \BibitemOpen
  \bibinfo {note} {With these notations, it is manifest that summing over $i$
  recovers Eq.~\protect \textup {\hbox {\mathsurround \z@ \protect \normalfont
  (\ignorespaces \ref {eq:diffusion_total}\unskip \@@italiccorr )}}, with $\rho
  ^\psi (\psi )=\DOTSB \sum@ \slimits@ _i \rho ^\psi _i(\psi )$.}\BibitemShut
  {Stop}%
\bibitem [{Note3()}]{Note3}%
  \BibitemOpen
  \bibinfo {note} {Note that the theory developed in this paper only describes
  particles trapped inside a box, whereas Refs.~\cite
  {Angelani2009,DiLeonardo2010} are concerned with active particles swimming
  outside the gear. However, Ref.~\cite {Sokolov2010} demonstrates that both
  configurations lead to a torque on the gear using very similar shapes,
  therefore the question of how to design the gear may be discussed from either
  point of view.}\BibitemShut {Stop}%
\bibitem [{Note4()}]{Note4}%
  \BibitemOpen
  \bibinfo {note} {Technically, $D$ is never reached. However, the particle
  gets within any reasonably small distance of $D$ within a few relaxation
  times.}\BibitemShut {Stop}%
\end{thebibliography}%

\end{document}